\documentclass[11pt,a4paper]{article}
\usepackage[dvips]{graphicx}
\usepackage{epsfig}
\usepackage{amssymb}
\usepackage{amsmath}
\usepackage{amsfonts}
\usepackage{amsthm}
\usepackage{amscd}
\usepackage{amsbsy}
\usepackage{bm}
\usepackage{url}
\usepackage{layout}
\usepackage{cite}
\usepackage{epstopdf}
\usepackage{color}
\usepackage[colorlinks=true, allcolors=blue, breaklinks=true]{hyperref}
\usepackage[utf8]{inputenc}
\usepackage[T1]{fontenc}
\usepackage{marvosym}
\usepackage{breakcites}
\usepackage{fourier}
\usepackage{apalike}
\usepackage{vmargin}
\setmarginsrb{2cm}{2cm}{2cm}{2cm}{0mm}{0mm}{0mm}{10mm}
\usepackage{titling}
\newtheorem{theorem}{Theorem}

\newtheorem{definition}[theorem]{Definition}

\title{\bfseries The nonlinear Schr\"odinger equation:\\ A mathematical model with its wide-ranging applications}
\author{\normalsize Natanael Karjanto\thanks{\Letter: \url{natanael@skku.edu}} \\ 
{\small Department of Mathematics, University College, Natural Science Campus}\\
{\small Sungkyunkwan University, Suwon 16419, Republic of Korea}}
\date{\vspace{-1cm}}

\begin{document}
\maketitle
\begin{abstract}
The nonlinear Schr\"odinger equation (NLSE) models the slowly varying envelope dynamics of a weakly nonlinear quasi-monochromatic wave packet in dispersive media. In the context of Bose-Einstein condensate (BEC), it is often referred to as the Gross-Pitaevskii equation (GPE). The NLSE is one example of integrable systems of a nonlinear partial differential equation (PDE) in $(1 + 1)$D and it possesses an infinite set of conservation laws. This nonlinear evolution equation arises in various physical settings and admits a wide range of applications, including but not limited to, surface gravity waves, superconductivity, nonlinear optics, and BEC. This chapter discusses not only the modeling aspect of the NLSE but also provides an overview of the applications in these four exciting research areas. The former features derivations of the NLSE heuristically and by employing the method of multiple-scale from other mathematical models as governing equations. Depending on how the variables are interpreted physically, the resulting NLSE can model a different dynamics of the wave packet. Furthermore, depending on the adopted assumptions and the chosen governing equations, each approach may provide different values for the corresponding dispersive and nonlinear coefficients.
\end{abstract}

\tableofcontents

\section{Introduction}

A lot of physical and natural phenomena can be modeled mathematically using partial differential equations (PDEs). In the realm of linear theory, solutions of PDEs obey the principle of linear superposition, and in some cases, they possess explicit analytical expressions. However, the laws of the nature are not always linear, and nonlinear PDEs often play an essential role in modeling these phenomena. Having applications in practically all areas of the natural sciences, the theory of PDEs, both linear and nonlinear, is one of the largest and most active areas of modern mathematics.

Historically, the subject of PDEs sprang as a study on the geometry of surfaces and tackled various problems in mechanics. It continues with the historically developed calculus of variations to bridge the theory of surfaces with the understanding of physical problems. Particularly, the study of wave propagation problems has, in turn, stimulated further developments in the general theory of nonlinear PDEs. Some examples of famous nonlinear PDEs amongst others are the nonlinear kinematic wave equation, the nonlinear Klein-Gordon equation, the Burgers equation, the Fischer equation, the Boussinesq equation, the Korteweg-de Vries equation (KdVE), the nonlinear Schr\"odinger equation (NLSE), the Benjamin-Ono equation, the Benjamin-Bona-Mahoney equation, the Kadomtsev-Petviashvili equation, the Davey-Stewartson equation, and the Camassa-Holm equation~\cite{Debnath12}.

This chapter specifically deals with the NLSE. In particular, we will provide an overview on modeling and application aspects of the model in $(1 + 1)$D, the simplest domain of space and time variables. The nonlinear term in the NLSE that we are discussing is in the form of the cubic power and thus the equation is also sometimes called the cubic Schr\"odinger equation (CSE). In the context of Bose-Einstein condensation (BEC), the NLSE is known as the Gross-Pitaevskii equation (GPE). The NLSE models an evolution equation for slowly varying envelope dynamics of a weakly nonlinear quasi-monochromatic wave packet in dispersive media. It possesses an infinite set of conservation laws and the model is a completely integrable system by the inverse scattering transform~\cite{Ablowitz74,Ablowitz81,Ablowitz91}. Indeed, having fundamental knowledge on the NLSE is essential in understanding the general theory of nonlinear dispersive waves.

In the absence of the nonlinear term, the NLSE reduces to the well-known Schr\"odinger equation, a linear PDE that governs a wave function characterizing the state of a quantum-mechanical system. The equation is named after the Austrian physicist Erwin Schr\"odinger who developed a number of fundamental results in the field of quantum theory~\cite{Schrodinger26}. The Schr\"odinger equation can be derived using the mathematical formulation of quantum mechanics in terms of operators in Hilbert space introduced by~\cite{Dirac30} and~\cite{von32}, known as the Dirac-von Neumann axioms. The readers who are interested in more detailed discussion on the Schr\"odinger equation may consult any textbook in quantum mechanics, such as~\cite{Grif18,Phil03} or~\cite{Shankar94}. The discussion in this chapter will focus more on the NLSE instead of the (linear) Schr\"odinger equation.

The NLSE arises in various physical settings in fluid mechanics and hydrodynamics describing the evolution of surface gravity water waves. It also has extensive applications in nonlinear optics, plasma physics and magnetohydrodynamics, solid state physics characterizing the propagation of a heat pulse in a solid, superconductivity describing solitary waves propagation in piezoelectric semiconductors, condensed matter such as BEC, and even mathematical finance. In this chapter, we will focus only on four of these applications: water waves, superconductivity, nonlinear optics and BEC.

This chapter is organized as follows. After this introduction, Section~\ref{hydrodynamics} discusses NLSE modeling and application in hydrodynamics. We derive both linear and nonlinear Schr\"odinger equations heuristically. By implementing the method of multiple-scale~\cite{Kevorkian12,Kevorkian13,Nayfeh08}, it follows with the derivation of the temporal and the spatial NLSEs. An overview of applications in surface water waves will be presented. Section~\ref{superconductivity} covers the NLSE derivation from the nonlinear Klein-Gordon equation using the same technique as in Section~\ref{hydrodynamics}. We also consider applications of the sine-Gordon models in superconductivity. Section~\ref{nonlinearoptics} reviews modeling and applications of the NLSE in the field of nonlinear optics. We adopt Maxwell's and Helmholtz' equations as a starting point for the derivation of the NLSE. Section~\ref{BEcondensate} describes the derivation of the GPE by means of the mean-field theory and its applications in the emerging field of BEC. Finally, Section~\ref{conclusion} provides concluding remarks to our discussion. 

\section{Hydrodynamics}  \label{hydrodynamics}

\subsection{Heuristic derivation for the NLSE}

The following heuristic derivation for the NLSE is based on~\cite{Debnath94,Dingemans97}.
For a linear dispersive wave equation, we can express its general solution in terms of a Fourier transform representation. 
Suppose we have the following linear equation governing the evolution of the surface elevation $\eta(x,t)$:
\begin{equation}
\partial_{t} \eta + i \Omega(-i\partial_{x}) \eta = 0,
\end{equation}
then its general solution $\eta(x,t)$ expressed by the Fourier representation is given by
\begin{equation}
\eta(x,t) = \frac{1}{2\pi} \int_{-\infty}^{\infty} F(\zeta) \, e^{i(k x - \omega t)} \, \mathrm{d}\zeta.  \label{generaleta}
\end{equation}
Here, we can replace the variable $\zeta$ either with wavenumber $k$ or with frequency $\omega$, and both are related by the linear dispersion relationship $\omega = \Omega(k)$ or $k = K(\omega)$, where $K = \Omega^{-1}$. The spectrum function $F(\zeta)$ will be determined from a given initial or boundary condition.
For an initial value problem (IVP), $F(k)$ is the Fourier transform of the initial condition $\eta(x,0)$.
Correspondingly, the Fourier transform of the initial signal $\eta(0,t)$ is given by $F(\omega)$ in the case of a boundary value problem (BVP).
They are given as follows:
\begin{align}
F(k)      &= \int_{-\infty}^{\infty} \eta(x,0) \, e^{-i k x}     \, \mathrm{d}x \\
F(\omega) &= \int_{-\infty}^{\infty} \eta(0,t) \, e^{ i \omega t} \, \mathrm{d}t.
\end{align}
We adopt an assumption of a slowly modulated wave as it propagates in a dispersive medium, and hence $F(k)$ and $F(\omega)$ are narrow-banded spectra around $k_0$ and $\omega_0$, respectively.
The linear dispersion relationship can be expressed in its Taylor-expansion series about the basic state wavenumber $k_0$ and frequency $\omega_0$, written as follows:
\begin{align}
\Omega(k) &= \omega_0 + \Omega'(k_0)(k - k_0) + \frac{1}{2!} \Omega''(k_0) (k - k_0)^2 + \dots \\
K(\omega) &= k_{0} + K'(\omega_{0})(\omega - \omega_{0}) + \frac{1}{2!}K''(\omega_{0})(\omega - \omega_{0})^{2} + \dots.
\end{align}
Therefore, we can rewrite~\eqref{generaleta} as $\eta(x,t) = A(\xi,\tau) e^{i(k_{0}x - \omega_{0}t)}$, where $A(\xi,\tau)$ is the corresponding complex-valued amplitude of the wave packet $\eta(x,t)$, written in two different versions:
\begin{align}
A(\xi_1,\tau_1) &= \frac{1}{2\pi} \int_{-\infty}^{\infty} F(k_0 + \kappa) \, e^{i(\xi_1 - \Omega_{\textmd{res}}(k_0)\tau_1/\kappa^{2})} \,\mathrm{d} \kappa, \label{complexamplitude1} \\
A(\xi_2,\tau_2) &= \frac{1}{2\pi} \int_{-\infty}^{\infty} F(\omega_{0} + \nu) \, e^{-i(\tau_2 - K_{\textmd{res}}(\omega_0)\xi_2/\nu^{2})} \, \mathrm{d} \nu. \label{complexamplitude2}
\end{align}
Here, $\kappa = k - k_0 = \mathcal{O}(\epsilon)$, $\xi_1 = \kappa(x - \Omega'(k_0) t)$, $\tau_1 = \kappa^2 t$,
$\nu = \omega - \omega_{0} = \mathcal{O}(\epsilon)$, $\xi_2 = \nu^{2} x$, $\tau = \nu(t - K'(\omega_{0})x)$, and $0 < \epsilon \ll 1$ is a small positive parameter.
The residual terms appearing in the exponential term read
\begin{align*}
\Omega_{\textmd{res}}(k_0) &= \Omega(k) - [\omega_0 + \Omega'(k_0)\kappa]  = \kappa^{2}\left(\frac{1}{2!}\Omega''(k_0) + \frac{1}{3!}\Omega'''(k_0)\kappa  + \dots \right) \\
K_{\textmd{res}}(\omega_0) &= K(\omega) - [k_{0} + K'(\omega_{0})\nu]  = \nu^{2}\left(\frac{1}{2!}K''(\omega_{0}) + \frac{1}{3!}K'''(\omega_{0})\nu  + \dots \right).
\end{align*}
From the complex-amplitude representations~\eqref{complexamplitude1} and~\eqref{complexamplitude2}, it follows that $\kappa$ (respectively $\nu$) are associated with the differential operator $i\partial_{\xi}$ (respectively $-i \partial_{\tau}$) and thus, $\kappa^2 = - \partial_{\xi}^2$ (respectively $\nu^{2} = - \partial_{\tau}^{2}$). Thus, the complex-valued amplitude $A$ satisfies
\begin{align}
\partial_{\tau}A + i \Omega_{\textmd{res}}( i\partial_{\xi})  A &= 0   \label{LSKres1} \\
\partial_{\xi}A  + i K_{\textmd{res}}     (-i\partial_{\tau}) A &= 0.  \label{LSKres2} 
\end{align}
For a narrow-banded spectrum, the equations~\eqref{LSKres1} and~\eqref{LSKres2} reduce to approximate equations called the temporal and the spatial ``linear Schr\"{o}dinger'' equations by approximating $\Omega_{\textmd{res}}$ and $K_{\textmd{res}}$ in their lowest-order terms, respectively
\begin{align}
i\partial_{\tau}A + \beta_1 \partial_{\xi}^{2}  A &= 0,  \label{linearSchrodinger1} \\
i\partial_{\xi} A + \beta_2 \partial_{\tau}^{2} A &= 0.  \label{linearSchrodinger2}
\end{align}
Here, the dispersion coefficients $\beta_1 = \frac{1}{2}\Omega''(k_0)$ and $\beta_2 = -\frac{1}{2} K''(\omega_{0}) = \frac{1}{2}\frac{\Omega''(k_{0})} {[\Omega'(k_{0})]^{3}}$.

Before proceeding to derive the NLSE heuristically, we make a quick note on the local existence and uniqueness for an IVP of the temporal linear Schr\"{o}dinger equation (LSE)~\eqref{linearSchrodinger1}. Physically, this evolution equation exhibits a dispersion phenomenon. It means that the group velocity $\Omega'(k)$ depends on the wavenumber $k$ and waves with different frequencies travel at different speed. Consequently, a traveling and localized wave packet will dissolve. 
\begin{definition}[Fundamental solution]
The function
\begin{equation}
\Psi(\xi,\tau) := \frac{1}{\sqrt{4 \pi i \beta_1 \tau}} e^{\frac{i|\xi|^2}{4 \beta_1 \tau}}, \qquad \xi \in \mathbb{R}, \quad \tau \neq 0.  \label{funso}
\end{equation}
is called the {\upshape fundamental solution} of the LSE~\eqref{linearSchrodinger1}.
\end{definition}
\begin{theorem}
The corresponding initial value problem LSE~\eqref{linearSchrodinger1} with an initial condition $A(\xi,0)$ admits an exact solution using the fundamental solution~\eqref{funso}, given explicitly as follows:
\begin{equation}
A(\xi,\tau) = \frac{1}{\sqrt{4\pi i \beta_1 \tau}} \int_{-\infty}^{\infty} e^{-\frac{(\xi - \zeta)^2}{4i \beta_1 \tau}} A(\zeta,0) \, \mathrm{d} \zeta, \qquad \xi \in \mathbb{R}, \quad \tau > 0.
\end{equation}
\end{theorem}
The proof of this theorem can be found in~\cite{Evans10,Fibich15} using the Fourier transform method and we have adopted the following convention for the Fourier transform definition in this chapter. 
\begin{definition}[Fourier transform]
Let $f$ be a defined function on $\xi \in \mathbb{R}$, then the {\upshape Fourier transform} of $f$ and its {\upshape inverse Fourier transform} are given as follows:
\begin{align}
\hat{f}(\kappa) = {\cal F}\left\{ f(\xi) \right\} &= \frac{1}{\sqrt{2 \pi}} \int_{-\infty}^{\infty} f(\xi) e^{-i \kappa \xi} \, \mathrm{d}\xi \\
f(\xi) = {\cal F}^{-1} \left\{ \hat{f}(\kappa) \right\} &= \frac{1}{\sqrt{2 \pi}} \int_{-\infty}^{\infty} \hat{f}(\kappa) e^{i \kappa \xi} \, \mathrm{d}\kappa.
\end{align} 
\end{definition}
\begin{proof}
Let $\hat{A}(\kappa,\tau) = {\cal F}\{ A(\xi,\tau) \}$ be the Fourier transform of the complex-valued amplitude $A(\xi,\tau)$.
Taking the Fourier transform of the LSE~\eqref{linearSchrodinger1} and its initial condition yields the following, respectively 
\begin{equation}
i \partial_{\tau} \hat{A} - \beta_1 \kappa^2 \hat{A} = 0, \qquad \qquad \text{and} \qquad \qquad \hat{A}(\kappa, 0).
\end{equation}
The solution to this ODE is 
\begin{equation}
\hat{A}(\kappa, \tau) = \hat{A}(\kappa,0) e^{-i \beta_1 \kappa^2 \tau}.  \label{invFTA}
\end{equation}
Taking the inverse of the Fourier transform~\eqref{invFTA}, we obtain the solution in convolution form
\begin{equation}
A(\xi,\tau) = \frac{1}{\sqrt{2\pi}} A(\xi,0) \ast {\cal F}^{-1} \left\{e^{-i \beta_1 \kappa^2 \tau} \right\}.
\end{equation}
Using the fact that the Fourier transform of a Gaussian-shape function is also Gaussian-shape profile
\begin{equation}
{\cal F} \left\{e^{-\frac{p}{2} \xi^2} \right\} = \frac{1}{\sqrt{p}} e^{-\frac{\kappa^2}{2p}}, \qquad \text{and} \qquad
{\cal F}^{-1} \left\{e^{-i \beta_1 \kappa^2 \tau} \right\} = \frac{1}{\sqrt{2 i \beta_1 \tau}} e^{- \frac{\xi^2}{4 i \beta_1 \tau}},
\end{equation}
with $p = 1/(2i\beta_1 \tau)$, we then obtain the desired expression
\begin{equation}
A(\xi,\tau) = \frac{1}{\sqrt{4\pi i \beta_1 \tau}} \int_{-\infty}^{\infty} e^{-\frac{(\xi - \zeta)^2}{4i \beta_1 \tau}} A(\zeta,0) \, \mathrm{d} \zeta, \qquad \xi \in \mathbb{R}, \quad \tau > 0.
\end{equation}
\end{proof}

This solution does make sense for all positive time $t > 0$ provided that the initial condition $A(\xi,0) \in L^{1}(\mathbb{R})$. From this expression, we obtain the following basic $L^{\infty}$-estimate~\cite{Strauss89,Cazenave96a,Teschl14}
\begin{align}
\left\|A(\xi,\tau) \right\|_{L^\infty(\mathbb{R})} = \sup_{\xi \in \mathbb{R}} \left| A(\xi,\tau) \right| &\leq \frac{1}{\sqrt{4 \pi \beta_1 \tau}} \int_{-\infty}^{\infty}  \left|A(\xi,0) \right| \, \mathrm{d}\xi \nonumber \\
&= \frac{1}{\sqrt{4 \pi \beta_1 \tau}} \left\| A(\xi,0) \right\|_{L^1(\mathbb{R})}.
\end{align}
This indicates that any solution with spatially localized initial conditions will decay uniformly toward zero with a rate of $\sqrt{\tau}$~\cite{Schneider11}.
Furthermore, using Plancherel's theorem, if the initial condition $A(\xi,0) \in L^1(\mathbb{R}) \cap L^2(\mathbb{R})$, then the $L^2$-norm is preserved
\begin{equation}
\left \|A(\xi,\tau) \right\|_{L^2(\mathbb{R})} = \left\| A(\xi,0) \right\|_{L^2(\mathbb{R})}, \qquad \qquad \forall \quad t > 0.
\end{equation}

In the following, we will derive the NLSE from the nonlinear dispersion relationship that includes a slowly-varying real-valued amplitude $a(x,t) \sim \mathcal{O}(\epsilon)$:
\begin{equation}
\omega = \Omega(k, a^2) \qquad \qquad \text{or} \qquad \qquad k = K(\omega, a^2).
\end{equation}
We perform a Taylor series expansion about the basic state wavenumber $k = k_0$ (or the basic state frequency $\omega = \omega_0$) and the zero amplitude $|a|^2 = 0$
\begin{align}
\omega &= \Omega(k_0,0) + \frac{\partial \Omega}{\partial k}(k_0,0) (k - k_0) + \frac{\partial^2 \Omega}{\partial |a|^2}(k_0,0) |a|^2 + \frac{1}{2} \frac{\partial^2 \Omega}{\partial k^2}(k_0,0) (k - k_0)^2 \nonumber \\
& \quad + \frac{\partial^2 \Omega}{\partial k \partial|a|^2}(k_0,0)(k - k_0)|a|^2 + \frac{1}{2} \frac{\partial^2 \Omega}{\partial|a|^4}(k_0,0)|a|^4 + \dots
\end{align}
Rewriting and considering only the essential terms, we obtain
\begin{equation}
\nu - \Omega'(k_0) \kappa - \frac{1}{2} \Omega''(k_0) \kappa^2 - \epsilon^2 \frac{\partial^2 \Omega}{\partial |A|^2}(k_0,0) |A|^2 - \dots = 0.
\end{equation}
Associating $\nu$ and $\kappa$ with differential operators $i\partial_{t}$ and $-i\partial_{x}$, and acting on the complex-valued amplitude $A$, it yields
\begin{equation}
i(\partial_{t} + \Omega'(k_0) \partial_{x}) A + \frac{1}{2} \Omega''(k_0) \partial_{x}^2 A - \epsilon^2 \frac{\partial^2 \Omega}{\partial |a|^2}(k_0,0) |A|^2 A = 0.
\end{equation}
Introducing the slowly-moving coordinate $\xi = \epsilon(x - \Omega'(k_0)t)$ and slower time variable $\tau = \epsilon^2 t$, we obtain the temporal NLSE
\begin{equation}
i \partial_{\tau} A + \beta_1 \partial_{\xi}^2 A + \gamma_1 |A|^2 A = 0 \label{temporalNLS1}
\end{equation}
where the dispersion and the nonlinear coefficients are respectively given by
\begin{equation}
\beta_1 = \frac{1}{2} \Omega''(k_0) \qquad \qquad \text{and} \qquad \qquad
\gamma_1 = -\frac{\partial^2 \Omega}{\partial |A|^2}(k_0,0).
\end{equation}

For the temporal NLSE~\eqref{temporalNLS1}, the following result on the local existence and uniqueness solution for an initial value problem in Sobolev spaces is known in the literature. See, for instance,~\cite{Schneider11,Kramer13} for a detailed proof of the following theorem. 
\begin{theorem}
Let $m \geq 1$, let $H^{m}(\mathbb{R})$ be the Sobolev space, the space of $m$ times weakly differentiable functions $u: \mathbb{R} \rightarrow \mathbb{R}$ with derivatives in $L^2$-space for $j \in \{0,1,\dots, m \}$. Let the space $H^m$ be equipped with the norm 
\begin{equation}
\| u \|_{H^m} = \max_{j \in \{0,1,\dots, m\}} \| \partial_\xi^j u \|_{L^2}.
\end{equation}
Let $A_0 \in H^m(\mathbb{R})$ be a complex-valued function. Then there exists a time $\tau_0 = \tau_0\left(\| A_0 \|_{H^m} \right) > 0$ and a unique solution $A \in \mathbb{C}\left( \left[0,\tau_0 \right], H^m \right)$ of the temporal NLSE~\eqref{temporalNLS1} with the initial condition $A_0$.
\end{theorem}
The readers who are interested in the well-posedness of the Cauchy problem and the long-time behavior of the corresponding global solutions to the NLSE may consult~\cite{Strauss89,Cazenave96a,Cazenave96b,Ginibre97,Bourgain99}.

Similarly, writing out the Taylor series expansion in two variables for $k = K(\omega,a^2)$ and retaining only the essential terms, we have
\begin{align*}
k &= K(\omega_0,0) + \frac{\partial K}{\partial \omega}(\omega_0,0) (\omega - \omega_0) + \frac{1}{2} \frac{\partial^2 K}{\partial \omega^2}(\omega_0,0) (\omega - \omega_0)^2 + \frac{\partial^2 \Omega}{\partial |a|^2}(\omega_0,0) |a|^2 + \dots
\end{align*}
Rearranging the terms to the left-hand side yields
\begin{equation}
\kappa - K'(\omega_0) \nu - \frac{1}{2} K''(\omega_0) \nu^2 - \frac{\partial^2 \Omega}{\partial |a|^2}(\omega_0,0) |a|^2 - \dots = 0.
\end{equation}
Corresponding the parameters $\kappa$ and $\nu$ with the differential operators $-i\partial_{x}$ and $i\partial_{t}$, respectively, and acting on the complex-valued amplitude $A$, we obtain
\begin{equation}
i \left(\partial_{x} + K'(\omega_0) \partial_{t}\right) A - \frac{1}{2} K''(\omega_0) \partial_{t}^2 A + \epsilon \frac{\partial^2 \Omega}{\partial |A|^2}(\omega_0,0) |A|^2 A = 0.
\end{equation}
Introducing the slowly-moving variables $\xi = \epsilon^2 x$ and $\tau = \epsilon(t - K'(\omega_0) x)$, we obtain the spatial NLSE
\begin{equation}
i \partial_{\xi} A + \beta_2 \partial_{\tau}^2 A + \gamma_2 |A|^2 A = 0
\end{equation}
where the dispersion and the nonlinear coefficients are respectively given by
\begin{equation}
\beta_2 = - \frac{1}{2} K''(\omega_0) \qquad \qquad \text{and} \qquad \qquad
\gamma_2 = \frac{\partial^2 \Omega}{\partial |A|^2}(\omega_0,0).
\end{equation}

\subsection{Derivation of the temporal NLSE}  \label{SubsectemporalNLSE}

The following derivation, as well as the derivation for the spatial NLSE in the next subsection, follow the argument presented by~\cite{vanGroesen98,Cahyono02,Karjanto06}.
Consider a KdV type of equation with an exact dispersion relationship property, given as follows
\begin{equation}
\partial_{t}\eta + i \Omega(-i\partial_{x})\eta + c \,\eta \partial_{x}\eta = 0. 		\label{KdVeqn}
\end{equation}
Here, $c \in \mathbb{R}  $ is a nonlinear coefficient for the KdV equation~\eqref{KdVeqn}. We will observe that it also contributes to the nonlinear coefficient for the NLSE. 
The function $\Omega$ acts as both a differential operator and a dispersion relationship. 

We seek a solution for the surface wave elevation $\eta(x,t)$ in the form of a wave packet, or a wave group.
This wave packet consists of a superposition of the first-order harmonic wave, the second-order double harmonic wave, and the second-order non-harmonic long wave, explicitly given as follows:
\begin{equation}
\eta(x,t) = \epsilon A(\xi,\tau)e^{i\theta} + \epsilon^{2}[B(\xi,\tau)e^{2i\theta} + C(\xi,\tau)] + \textmd{c.c.},		   \label{etaABC}
\end{equation}
where $0 < \epsilon \ll 1$ is a small positive parameter as used commonly in perturbation theory. The term in the exponent is $\theta(x,t) = k_{0}x - \omega_{0}t$, where $k_0$ and $\omega_0$ are related by the linear dispersion relation. The functions $A(\xi,\tau)$, $B(\xi,\tau)$, and $C(\xi,\tau)$ are complex-valued wave packet envelopes and they are allowed to vary slowly in the slower-moving frame of reference, where the spatial and temporal variables are given by $\xi = \epsilon(x - \Omega'(k_{0})t)$ and $\tau =
\epsilon^{2}t$, respectively. As usual, c.c. denotes the complex conjugation of the preceding terms.

Substituting the Ansatz~\eqref{etaABC} into the KdV equation~\eqref{KdVeqn} yields the residue $R(x,t)$ that can be expressed in the following form:
\begin{equation}
R(x,t) = \sum_{n,m} \epsilon^{n} R_{nm} e^{i \, m \theta} + \textmd{c.c.}, 			\label{Rnm}
\end{equation}
where $n \geq 1$, $m \geq 0$, and where the coefficients $R_{nm}$ contain expressions in $A(\xi,\tau)$, $B(\xi,\tau)$, and $C(\xi,\tau)$ and their partial derivatives. 
All coefficients of $R(x,t)$ must vanish in order to satisfy the KdV equation~\eqref{KdVeqn}. 
We obtain the vanishing of the first order residue coefficients $R_{1m} = 0$, $m \geq 0$ when $k_{0}$ and $\omega_{0}$ are related by the linear dispersion relationship. 
The second order residue coefficients read
\begin{align*}
  R_{20} &= 0; \qquad  R_{21}  =  0; \qquad  R_{2m} = 0, \quad m \geq 3; \\
  R_{22} &= i([\Omega(2k_{0}) - 2\omega_{0}]B + c\,k_{0}A^{2}).
\end{align*}
Vanishing of $R_{22}$ leads to an expression of $B(\xi,\tau)$ as a function of $A(\xi,\tau)$:
\begin{equation}
B(\xi,\tau) = \frac{c\,k_{0}A^{2}(\xi,\tau)}{2\omega_{0} - \Omega(2k_{0})}.		  \label{BR22}
\end{equation}
The third order residue coefficients read
\begin{align*}
  R_{30} &= [\Omega'(0) - \Omega'(k_{0})]\partial_{\xi}C + \frac{1}{2}c \, \partial_{\xi}|A|^{2};\\
  R_{31} &= \partial_{\tau}A - \frac{1}{2}i\Omega''(k_{0}) \partial_{\xi}^{2}A + i\, c k_{0} (A^{\ast} B + AC + A C^{\ast}).
\end{align*}
Requiring $R_{30}$ to vanish yields an expression for $C(\xi,\tau)$ as a function of $A(\xi,\tau)$ and a $\tau$-dependent constant of integration $\alpha_{T}$, given as follows:
\begin{equation}
C(\xi,\tau) = \frac{1}{2} \frac{c\, |A(\xi,\tau)|^{2}}{\Omega'(k_{0}) - \Omega'(0)} + \alpha_{T}(\tau). 		\label{CR30}
\end{equation}
To prevent a resonance, $R_{31}$ has to vanish as well, and this leads to an evolution equation for $A(\xi,\tau)$:
\begin{equation}
\partial_{\tau}A + i \beta_1 \partial_{\xi}^{2} A +  i \gamma_1 |A|^{2}A  +  2 i k_{0} c\, \textmd{Re}[\alpha_{T}(\tau)] A = 0.
\end{equation}
A similar assumption of unidirectional wave propagation, applying the ``gauge transformation'' by multiplying the evolution equation by $e^{2ik_{0}c \int \textmd{Re}[\alpha_{T}(\tau)] d\tau}$~\cite{Mei83}, we obtain the temporal NLSE for $A(\xi,\tau)$:
\begin{equation}
\partial_{\tau}A + i \beta_1 \partial_{\xi}^{2} A +  i \gamma_1 |A|^{2}A = 0,		  \label{temporalNLS}
\end{equation}
where the dispersion and the nonlinear coefficients are respectively given by
\begin{align}
\beta_1  &= -\frac{1}{2}\Omega''(k_{0}) \\
\gamma_1 &=  k_{0}c^{2} \left(\frac{1}{\Omega'(k_{0}) - \Omega'(0)} + \frac{k_{0}}{2\omega_{0} - \Omega(2k_{0})} \right).
\end{align}
This temporal NLSE as well as the spatial NLSE derived in the following subsection are valid for intermediate-water wave models.

In the following, we will derive the corresponding ``energy equation'' and ``nonlinear dispersion relationship'' for the temporal NLSE.
Write the complex-valued amplitude $A(\xi,\tau)$ in the physical, faster-moving variables as $A_1(x,t) = \epsilon A(\xi,\tau)$ using the relationship above $\xi = \epsilon(x - \Omega'(k_{0})t)$ and $\tau = \epsilon^{2}t$.
The temporal NLSE in the physical variables is given by
\begin{equation}
\partial_{t} A_1 + \Omega'(k_{0}) \partial_{x} A_1 + i \beta_1 \partial_{x}^{2} A_1 + i \gamma_1 |A_1|^{2} A_1 = 0.			\label{temporalNLSphysical}
\end{equation}
Apply the Madelung transformation by writing the complex-valued amplitude $A_1(x,t)$ in its polar form $A_1(x,t) = a(x,t) e^{i \phi(x,t)}$, where the amplitude $a(x,t)$ and the phase $\phi(x,t)$ are both real-valued functions~\cite{Madelung27}. Upon substitution to~\eqref{temporalNLSphysical}, removing the factor $e^{i\phi(x,t)}$ and separating the real and the imaginary parts, it yields the following coupled phase-amplitude equations
\begin{align}
\left\{
\begin{array}{ll}
\displaystyle{\partial_{t}a    + \Omega'(k_{0}) \partial_{x} a   - \beta_1 \left(a \partial_{x}^{2} \phi + 2 \partial_{x} a \partial_{x} \phi \right)}          &= 0 \\
\displaystyle{\partial_{t}\phi + \Omega'(k_{0}) \partial_{x}\phi + \beta_1 \left(\frac{\partial_{x}^{2}a}{a} - (\partial_{x}\phi)^{2} \right) + \gamma_1 a^{2}} &= 0.
\end{array}
\right. \label{phaseamplitudeeqns1}
\end{align}
Expressing the wavenumber $k(x,t)$ and the frequency $\omega(x,t)$ in the following form 
\begin{equation}
  k(x,t)      = k_0 + \kappa = k_{0}      + \partial_{x} \phi \qquad \qquad
  \omega(x,t) = \omega_0 + \nu = \omega_{0} - \partial_{t} \phi.
\end{equation}
The local wavenumber $\kappa = \partial_{x} \phi$ and the local frequency $\nu = -\partial_{t} \phi$ act as modulational quantities.
Using these quantities, the phase-amplitude equations can be written in a more compact form. 
The amplitude and the phase equations are the first and the second expressions in~\eqref{phaseamplitudeeqns1}, respectively.
Expressing $\partial_{x} \phi$ in terms of the local wavenumber $k(x,t)$ and multiplying the amplitude equation with $a(x,t)$, we obtain
\begin{equation}
\frac{1}{2} \partial_{t}(a^{2}) + \frac{1}{2} \partial_{x} \left\{ \left[\Omega'(k_{0}) + \Omega''(k_{0})(k - k_{0}) \right] a^{2} \right\} = 0.
\end{equation}
By noting the terms inside the square brackets on the second term above as a linear approximation for $\Omega'(k)$, we arrive at the conservation of energy equation for the temporal NLSE. It reads
\begin{equation}
\partial_{t}(a^{2}) + \partial_{x}[\Omega'(k) a^{2}] = 0.
\end{equation}
By expressing the local wavenumber $\partial_{x} \phi = k - k_0$ and the local frequency $-\partial_{t} \phi = \omega - \omega_0$, we obtain the following 
\begin{equation}
\omega - \left[\omega_{0} + \Omega'(k_{0})(k - k_{0}) + \frac{1}{2}\Omega''(k_{0})(k - k_{0})^{2} \right] = \beta_1 \frac{\partial_{x}^{2}a}{a} + \gamma_1 \, a^{2}.
\end{equation}
By taking the terms inside the square brackets as a quadratic approximation for $\Omega(k)$, the phase equation leads to the nonlinear dispersion relationship
\begin{equation}
\omega - \Omega(k) = \beta_1 \frac{\partial_{x}^{2}a}{a} + \gamma_1 a^{2}.		  \label{nondisreltem}
\end{equation}

\subsection{Derivation of the spatial NLSE} \label{spatialNLSE}

A similar technique using the method of multiple-scale can be applied to the KdVE with an exact dispersion relationship~\eqref{KdVeqn} to obtain the spatial NLSE.
The difference is in the choice of the slow-moving spatial and temporal variables, respectively chosen as $\xi = \epsilon^{2}x$ and $\tau = \epsilon(t - x/\Omega'(k_{0}))$.
Following an equivalent procedure as in the previous subsection, we obtain identical first-order and second-order residue coefficients
\begin{align*}
  R_{1m} &= 0, \quad m \geq 0 \\
  R_{20} &= 0; \qquad  R_{21}  =  0; \qquad  R_{2m} = 0, \quad m \geq 3; \\
  R_{22} &= i([\Omega(2k_{0}) - 2\omega_{0}]B + c\,k_{0}A^{2}).
\end{align*}
The third-order residue coefficients are given as follows:
\begin{align}
R_{30} &= \left(1 - \frac{\Omega'(0)}{\Omega'(k_{0})} \right) \partial_{\tau}C - \frac{1}{2} \frac{c}{\Omega'(k_{0})} \partial_{\tau} |A|^{2} \\
R_{31} &= \Omega'(k_{0})\partial_{\xi}A - \frac{1}{2}i  \frac{\Omega''(k_{0})}{[\Omega'(k_{0})]^{2}} \partial_{\tau}^{2}A + i\, c k_{0} (A^{\ast}B + AC + A C^{\ast}) \\
R_{32} &= -\frac{1}{\Omega'(k_{0})}\left(\Omega'(2k_{0})\partial_{\tau}B +  \frac{1}{2} c \partial_{\tau} A^{2} \right) \\
R_{33} &= 3ik_{0}c\,A B.
\end{align}
Requiring $R_{30}$ to vanish leads to an expression for $C(\xi,\tau)$ as a function of $A(\xi,\tau)$ and a $\xi$-dependent constant of integration $\alpha_{S}(\xi)$ for all $\theta(x,t)$:
\begin{equation}
C(\xi,\tau) = \frac{1}{2} \frac{c\, |A(\xi,\tau)|^{2}}{\Omega'(k_{0}) - \Omega'(0)} + \alpha_{S}(\xi).
\end{equation}
In order to prevent resonance, $R_{31}$ has to vanish which leads to a dynamic evolution equation for $A(\xi,\tau)$:
\begin{equation}
\partial_{\xi}A + i \beta_2 \partial_{\tau}^{2} A + i \gamma_2 |A|^{2}A + \frac{2ik_{0} c}{\Omega'(k_{0})} \textmd{Re}\left[\alpha_{S}(\xi) \right] A = 0.  \label{spatialNLSgauge}
\end{equation}

We can remove the term containing Re$[\alpha_{S}(\xi)]$ by applying the gauge transformation~\cite{Mei83}.
Multiply the evolution equation by $e^{\frac{2ik_{0}c}{\Omega'(k_{0})} \int \textmd{Re}[\alpha_{S}(\xi)] d\xi}$,
then the new complex amplitude $\tilde{A}(\xi,\tau) = e^{\frac{2ik_{0}c}{\Omega'(k_{0})} \int \textmd{Re}[\alpha_{S}(\xi)] d\xi} A(\xi,\tau)$ satisfies the spatial NLSE which, after dropping the tilde, can be written in the following form:
\begin{equation}
\partial_{\xi}A + i \beta_2 \partial_{\tau}^{2} A + i \gamma_2 |A|^{2}A = 0.    \label{spatialNLS}
\end{equation}
The dispersion coefficient $\beta$ and the nonlinear coefficient $\gamma$ are given as follows
\begin{align}
\beta_2  &= \beta(k_{0}) =  - \frac{1}{2} \frac{\Omega''(k_{0})}{[\Omega'(k_{0})]^{3}} \label{beta} \\
\gamma_2 &= \gamma(k_{0},c^2) = \frac{k_{0}c^{2}}{\Omega'(k_{0})} \left(\frac{1}{\Omega'(k_{0}) - \Omega'(0)} + \frac{k_{0}}{2\omega_{0} - \Omega(2k_{0})} \right).  \label{gamma}
\end{align}

We apply a similar approach as in the previous section to derive the corresponding energy equation and nonlinear dispersion relationship for the spatial NLSE.
Write $A_2(x,t) = \epsilon A(\xi,\tau)$, where $\xi = \epsilon^{2}x$ and $\tau = \epsilon(t - x/\Omega'(k_{0}))$.
The spatial NLSE in the physical variables is expressed as follows:
\begin{equation}
\partial_{x} A_2 + \frac{1}{\Omega'(k_{0})} \partial_{t} A_2 + i \beta_2 \partial_{t}^{2} A_2 + i \gamma_2 |A_2|^{2} A_2 = 0.
\label{spatialNLSphysical}
\end{equation}
Apply the Madelung transformation by writing $A_2(x,t)$ in its polar form $A_2(x,t) = a(x,t)e^{i\phi(x,t)}$, with $a(x,t)$ and $\phi(x,t)$ are real-valued quantities~\cite{Madelung27}.
After substituting to the spatial NLSE~\eqref{spatialNLSphysical}, removing the factor $e^{i\phi(x,t)}$, and collecting the real and the imaginary parts, we obtain the following coupled phase-amplitude equations in the original physical variables:
\begin{align}
\left\{
\begin{array}{ll}
\displaystyle{\partial_{x}a    + \frac{\partial_{t}a   }{\Omega'(k_{0})} - \beta_2 \left(a \partial_{t}^{2} \phi + 2 \partial_{t}a \partial_{t}\phi \right)}            &= 0 \\
\displaystyle{\partial_{x}\phi + \frac{\partial_{t}\phi}{\Omega'(k_{0})} + \beta_2 \left(\frac{\partial_{t}^{2}a}{a} - (\partial_{t}\phi)^{2} \right) + \gamma_2 a^{2}} &= 0.
\end{array}
\right. \label{phaseamplitudeeqns2}
\end{align}
Using the previous definition for wavenumber and frequency expressed in terms of modulated local wavenumber and local frequency, respectively, we can write these phase-amplitude equations in a more compact form.
We also adopt the linear dispersion relationship $\omega = \Omega(k)$ or $k = K(\omega)$, where $K = \Omega^{-1}$. From $k = K[\Omega(k)]$, we can derive the relationship between its derivatives, up to the second order given explicitly as follows:
\begin{equation}
  K'(\omega_{0}) = \frac{1}{\Omega'(k_{0})} \qquad \textmd{and} \qquad
  K''(\omega_{0})= -\frac{\Omega''(k_{0})}{[\Omega'(k_{0})]^{3}}.
\end{equation}
Expressing the local frequency $-\partial_{t}\phi = \omega - \omega_0$ and multiplying the amplitude equation with $a(x,t)$, we obtain
\begin{equation}
\frac{1}{2} \partial_{x}(a^{2}) + \frac{1}{2} \partial_{t} \left\{ \left[K'(\omega_{0}) + K''(\omega_{0})(\omega - \omega_{0}) \right] a^{2} \right\} = 0.
\end{equation}
Noting the terms in the square brackets as a linear approximation for $K'(\omega)$, we can write the amplitude equation as the energy equation
\begin{equation}
\partial_{x}(a^{2}) + \partial_{t}[K'(\omega) a^{2}] = 0.  \label{energyequation}
\end{equation}
By expressing the local wavenumber $\partial_{x} \phi = k - k_0$ and the local frequency $-\partial_{t} \phi = \omega - \omega_0$, we can write the phase equation as follows:
\begin{equation}
\left[k_{0} + K'(\omega_{0})(\omega - \omega_{0}) + \frac{1}{2}K''(\omega_{0})(\omega - \omega_{0})^{2} \right] - k = \beta_2 \frac{\partial_{t}^{2}a}{a} + \gamma_2 \, a^{2}.
\end{equation}
By considering the terms inside the square brackets as a quadratic approximation for $K(\omega)$, the phase equation leads to the nonlinear dispersion relationship
\begin{equation}
K(\omega) - k = \beta_2 \frac{\partial_{t}^{2}a}{a} + \gamma_2 \, a^{2}.   \label{nondisrelspa}
\end{equation}

The nonlinear dispersion relationship~\eqref{nondisreltem} or~\eqref{nondisrelspa} describes the relationship between wavenumber $k$ and frequency $\omega$ in the dispersion plane $(k,\omega)$. Since generally the right-hand side of ~\eqref{nondisreltem} or~\eqref{nondisrelspa} does not vanish, then any combination of $(k,\omega)$ does not always satisfy the linear dispersion relationship.

The ratio $\frac{\partial_{x}^2 a}{a}$ in~\eqref{nondisreltem} or $\frac{\partial_{t}^2 a}{a}$ in~\eqref{nondisrelspa} is coined as the ``Chu-Mei quotient'' by~\cite{Karjanto07a} when they discussed ``phase singularity'' and ``wavefront dislocation'' in surface gravity waves. The unboundedness of this quotient at the vanishing real-valued amplitude $a(x,t)$ is responsible for the occurrence of the phenomena. Several authors refer to this quotient as the ``Fornberg-Whitham term''~\cite{Infeld00}, referring to the paper written by~\cite{Fornberg78}. 
Even though the quotient has already appeared earlier in the literature in the context of modulated waves in nonlinear media~\cite{Karpman67,Karpman69}, it was~\cite{Chu70,Chu71} who introduced it for the first time when deriving the modulation equations of Whitham's theory for slowly varying Stokes' waves.

\subsection{Applications in surface gravity waves}

The temporal NLSE derived in Subsection~\ref{SubsectemporalNLSE} models absolute dynamics while the spatial NLSE derived in Subsection~\ref{spatialNLSE} models convective dynamics~\cite{Sulem99}. Together with an initial condition, the temporal NLSE constitutes an initial value problem. Given a wave packet profile at a particular time, the NLSE governs an evolution in time of the wave packet so that we could find out the shape of the wave packet at some time in the future.
Together with a boundary condition, the spatial NLSE composes a boundary value problem, or signaling problem. 
This model is suitable for a wave signal generation in a wave tank of a hydrodynamic laboratory.
By inputting an initial wave signal to a wavemaker, letting it propagate along the wave tank, we can measure the ``experimental'' wave signal at several frontal positions and compared it with the ``theoretical'' wave signal predicted by the spatial NLSE.

In particular, the spatial NLSE and its family of solitons on a non-vanishing background describing a nonlinear extension of modulational instability have been utilized as a model for deterministic freak wave generation in intermediate water depth at the high-speed wave basin of the Dutch Maritime Research Institute, the Netherlands (MARIN)~\cite{vanGroesen06,Karjanto06,Andonowati07}. The experimental results confirm an occurrence of phase singularity of the wave packet envelope at the location where the wave signals reach maximum amplitude, where the phenomenon and its related counterpart of wavefront dislocation have been predicted theoretically~\cite{Karjanto07a}. Even though the model does not quantitatively predict the signal and the spectrum evolution in accurate detail, it exhibits an extraordinary qualitative agreement. Our experimental results confirm similar behavior with other testings where the corresponding wave spectra demonstrate frequency downshift as the wave signals propagate along the wave tank~\cite{Karjanto10,Lake77}. The authors also coined the term ``Wessel curves'' indicating the evolution of the real-part and the imaginary-part of the complex-valued amplitude $A(\xi,\tau)$ exclusive of the oscillatory part contributed from the continuous-wave or the plane-wave solution.

Modulational instability, also known as sideband instability, is a well-known phenomenon in both fluid dynamics and nonlinear optics. In the context of hydrodynamics, it is known as the Benjamin-Feir instability where~\cite{Benjamin67,Benjamin67a} predicted the onset of the instability of Stokes wave trains in deep water. Under the theory of linear perturbation analysis, the plane-wave solution of the NLSE is modulationally unstable and its nonlinear extension is given by the ``Akhmediev-Eleonskii-Kulagin breather'', also known as the ``solitons on a non-vanishing background''~\cite{Akhmediev87,Akhmediev97,Ablowitz90,Karjanto09}. In the spectral domain, the effect of nonlinearity which reinforces periodic wave trains, leads to the generation of spectral sidebands and an eventual breakup of the waveform into a train of pulses. The breather's analytical expression for $\beta_1 = 1$ and $\gamma_1 =2$ in the temporal NLSE~\eqref{temporalNLS} can be written as follows:
\begin{equation}
A_\text{AEK}(\xi,\tau) = e^{2i\tau} \left(\frac{\nu^3 \cosh\left[\sigma (\tau - \tau_0) \right] + i \nu \sigma \sinh\left[\sigma (\tau - \tau_0)\right]}{2 \nu \cosh \left[\sigma (\tau - \tau_0)\right] - \sigma \cos \left[\nu (\xi - \xi_0) \right]} - 1 \right). 
\label{AEKbreather}
\end{equation}
The family of solitons on a non-vanishing background is defined for the parameter values $0 < \nu < 2$ and $\sigma = \nu \sqrt{4 - \nu^2}$. The soliton is a holomorphic function for $\sqrt{3} < \nu < 2$ and it reaches local maxima and local minima at $(\xi,\tau) = (\xi_0 + 2n\pi/\nu,\tau_0)$ and $(\xi,\tau) = (\xi_0 + [2n + 1]\pi/\nu, \tau_0)$, respectively, for $n \in \mathbb{Z}$.

Other NLSE solutions with a non-vanishing background type of soliton have been proposed for a hydrodynamic freak wave formation, called ``breather solutions''~\cite{Dysthe99}.
One family of solitons on a non-vanishing background is called by a new name: ``Kusnetsov-Ma breather''~\cite{Kibler12}, derived independently by~\cite{Kuznetsov77} and~\cite{Ma79}. For the corresponding dispersive and nonlinear coefficients in the temporal NLSE~\eqref{temporalNLS}, $\beta_1 = 1$ and $\gamma_1 = 2$, it is explicitly given as follows:
\begin{equation}
A_\text{KM}(\xi,\tau) = e^{2i\tau} \left(\frac{\mu^3 \cos \left[\rho (\tau - \tau_0) \right] + i \mu \rho \sin\left[\rho (\tau - \tau_0) \right]}{2 \mu \cos \left[\rho (\tau - \tau_0) \right] - \rho \cosh \left[\mu (\xi - \xi_0) \right]} + 1 \right)  \label{KMbreather}
\end{equation}
where $\rho = \mu \sqrt{4 + \mu^2}$, $\mu \in \mathbb{R}$.
Another exact solution is known as the ``Peregrine breather'' or the ``rational soliton'' solution~\cite{Peregrine83}. It is given as follows:
\begin{equation}
A_\text{PR}(\xi,\tau) = e^{2i\tau} \left(\frac{4(1 + 4i (\tau - \tau_0)}{1 + 16 (\tau - \tau_0)^2 + 4 (\xi - \xi_0)^2} - 1 \right).
\end{equation}
This solution can be obtained as a limiting case for both the Akhmediev-Eleonskii-Kulagin breather and the Kusnetsov-Ma breather when the parameters $\nu$ and $\mu$ approach zero~\cite{Karjanto07b}. The Peregrine breather has also been successfully generated experimentally as a rogue wave in a water wave tank~\cite{Chabchoub11}.

The NLSE has also been considered as a model to investigate the oceanic rogue wave formation caused by a nonlinear energy transfer in the open ocean, both deterministically and stochastically~\cite{Henderson99,Onorato01,Osborne01}. Extensive reports on progress in the physical mechanisms of oceanic rogue wave phenomenon are available~\cite{Kharif03,Dysthe08,Kharif09,Pelinovsky16}. A similar phenomenon has also been proposed, predicted, observed, and studied in other fields than hydrodynamics where the NLSE has been used as a mathematical model, including but not limited to, in optical rogue waves~\cite{Solli07}, atmospheric rogue waves~\cite{Stenflo10}, matter rogue waves in Bose-Einstein condensates~\cite{Bludov09}, and financial rogue waves~\cite{Ivancevic10,Yan10}.

\section{Superconductivity}  \label{superconductivity}

\subsection{NLSE derivation from the nonlinear Klein-Gordon equation}

Consider a line of pendula positioned very close together and they hang vertically under the influence of gravity.
There exists a horizontal torsion wire for which each pendulum can twist.
Let $u(x,t)$ be the twist angle of the pendulum at position $x$ and time $t$, then its motion can be modeled by a sine-Gordon equation
\begin{equation}
\partial^2_t u - a \partial^2 x u + b \sin u = 0, \qquad \qquad a, \; b \geq 0.
\end{equation}
In this model, the term $-b \sin u$ is an external force due to gravitational acceleration and the term $a \partial_x^2 u$ models the force caused by the effect of the twist.
Assume that one end of the pendulum chain is wiggled with a small amplitude motion with frequency $\omega$, then the term $\sin u$ can be approximated by its Maclaurin series about $u = 0$. Keeping only the first two terms, we obtain a nonlinear Klein-Gordon equation with a cubic nonlinearity.

The following derivation of the NLSE from a nonlinear Klein-Gordon equation follows the argument in~\cite{Sharma76,Newell85,Sulem99,Kramer13,Schneider11}.
Consider a cubic nonlinearity Klein-Gordon equation describing a model for a wave packet $u(x,t)$ that moves at a constant group velocity $c$, presented as the following  initial value problem (IVP):
\begin{align}
\partial_{t}^2 u - a \partial_{x}^2 u + b u - \lambda u^3 &= 0, \qquad \qquad a, \; b \geq 0, \quad \lambda \in \mathbb{R}   \label{KGE} \\
u(x,0) = u_0, \qquad \qquad \partial_{t}u(x,0) &= u_1.                                                                       \label{ICKGE}
\end{align}
An Ansatz for $u(x,t)$ is expressed as a perturbation series, where again $0 < \epsilon \ll 1$ is a small parameter
\begin{equation}
u(x,t) = \hat{u}_0 + \epsilon \hat{u}_1 + \epsilon^2 \hat{u}_2 + \dots,   \label{anzats}
\end{equation}
with $u_n(x,X,t,\tau_1,\tau_2)$, where $X = \epsilon x$, $\tau_1 = \epsilon t$, and $\tau_2 = \epsilon^2 t$ act as slower variables.
Substituting~\eqref{anzats} to the IVP~\eqref{KGE} and~\eqref{ICKGE} yields a series progressing in the order of $\epsilon$. The vanishing of the lowest-order term simply reduces to a linear Klein-Gordon IVP:
\begin{align}
\partial_{t}^2 \hat{u}_0 - a \partial_{x}^2 \hat{u}_0 + b \hat{u}_0 &= 0  \\
\hat{u}_0(x,X,0,0,0;\epsilon) &= u_0/\epsilon  \qquad \qquad \partial_{t}\hat{u}_0(x,X,0,0,0;\epsilon) = u_1/\epsilon.
\end{align}
We seek the solution in the form
\begin{equation}
\hat{u}_0(x,X,t,\tau_1,\tau_2) = A(X,\tau_1,\tau_2) e^{i(kx - \omega t)} + \text{c.c.}, \qquad A \in \mathbb{C}
\end{equation}
where c.c. denotes the complex conjugation of the preceding term and $(k,\omega)$ satisfies the dispersion relationship $\omega = \Omega(k) = \sqrt{a k^2 + b}$.
The vanishing of the first-order terms reads 
\begin{align}
\partial_{t}^2 \hat{u}_1 - a \partial_{x}^2 \hat{u}_1 + b \hat{u}_1 &= -2 \partial_{\tau_1} \partial_{t} \hat{u}_0 + 2 a \partial_{X} \partial_x \hat{u}_0 \nonumber \\
&= 2i \left(\omega \partial_{\tau_1} A + a k \partial_{X} A \right) e^{i(kx - \omega t)} + \text{c.c.} \\
\hat{u}_1(x,X,0,0,0;\epsilon) &= 0  \\
\partial_{t}\hat{u}_1(x,X,0,0,0;\epsilon) &= -\partial_{\tau_1} \hat{u}_0(x,X,0,0,0;\epsilon).
\end{align}
Since the right-hand side represents secular terms that would lead to unbounded growth in $\hat{u}_1$ over a long period of time, we need to eliminate them by taking $\omega \partial_{\tau_1} A + a k \partial_{X} A = 0$. This condition is equivalent to the group velocity $\Omega'(k) = a k/\Omega(k)$.
The solution of the linear Klein-Gordon equation for $\hat{u}_1$ is similar to the one for $\hat{u}_0$:
\begin{equation}
\hat{u}_1(x,X,t,\tau_1,\tau_2) = B(X,\tau_1,\tau_2) e^{i(kx - \omega t)} + \text{c.c.}, \qquad B \in \mathbb{C}.
\end{equation}
Collecting the second-order term and requiring it to vanish gives 
\begin{align}
\partial_{t}^2 \hat{u}_2 - \alpha \partial_{x}^2 \hat{u}_2 + \beta \hat{u}_2 
&= -2 \partial_{\tau_2} \partial_{t} \hat{u}_0 - \partial_{\tau_1}^2 \hat{u}_0 + a \partial_{X}^2 \hat{u}_0 + \lambda \hat{u}_0^3 + 2 \partial_{\tau_1} \partial_{t} \hat{u}_1  - 2a \partial_X \partial_{x} \hat{u}_1 \nonumber \\
&= \left\{ 2i\Omega(k) \partial_{\tau_2}A + (a - \Omega'(k)^2) \partial_{\xi}^2 A + 3\lambda |A|^2 A  \right. \nonumber \\ 
& \qquad \left. +   2i \left(\omega \partial_{\tau_1}B + a k \partial_{X} B \right) \right\} e^{i(kx - \omega t)} + \lambda A^3 e^{3i(kx - \omega t)} + \text{c.c.}\\
\hat{u}_2(x,X,0,0,0;\epsilon) &= 0  \\
\partial_{t}\hat{u}_2(x,X,0,0,0;\epsilon) &= -\partial_{\tau_1} \hat{u}_0(x,X,0,0,0;\epsilon) - \partial_{\tau_2} \hat{u}_0(x,X,0,0,0;\epsilon).
\end{align}
Here, we have used $\xi = \epsilon(x - \Omega'(k) t)$. Similar to previous method, we would like to remove secular terms by requiring $\hat{u}_2$ to be bounded and $B$ must satisfy $\omega \partial_{\tau_1} B + a k \partial_{X} B = 0$. For $A(\xi,\tau_2)$, it satisfies the temporal NLSE
\begin{equation}
i \partial_{\tau} A + \beta \partial_{\xi}^2 A + \gamma |A|^2 A = 0
\end{equation}
where we drop the subscript 2 from the variable $\tau_2$ and the dispersive and the nonlinear coefficients are given as follows, respectively
\begin{equation}
\beta = \frac{1}{2} \Omega''(k) \qquad \qquad \text{and} \qquad \qquad
\gamma = \frac{3}{2} \frac{\lambda}{\Omega(k)}.
\end{equation}
The additional term $\lambda A^3 e^{3i(kx - \omega t)} + \text{c.c.}$ is not a resonant term and thus, is not problematic since generally $\Omega(3k) \neq 3 \Omega(k)$.

\subsection{Applications of sine-Gordon model}

Both the nonlinear Klein-Gordon and the sine-Gordon equations have been studied analytically using the Zakharov-Shabat method~\cite{Zakharov72,Grundland92} and the sine-cosine and tanh methods~\cite{Wazwaz05}, as well as numerically using finite difference method~\cite{Strauss78,Jimenez90,Vu93,Duncan97,Li95} and using thin plate splines--radial basis functions~\cite{Dehghan09}.

These evolution equations find many applications in physical sciences and different nonlinear phenomena~\cite{Dodd82,Drazin89}.
Historically, the sine-Gordon equation arises from the field of differential geometry where it describes surfaces with a constant negative Gaussian curvature~\cite{Enneper70}.
The propagation of a crystal dislocation with sinusoidal periodicity is governed by the sine-Gordon equation~\cite{Frenkel39}. 
The behavior and interaction of elementary particles of mesons and baryons using an identical model has been proposed by~\cite{Perring62}.
The real-valued amplitude of wave packet envelope nonlinear evolution equations governing modulationally weakly unstable baroclinic shear flow can be transformed to the sine-Gordon equation~\cite{Gibbon79}. 

Another important application is in the field of superconductivity, which is known as Josephson effect across a Josephson junction.
The latter is a quantum mechanical device composed of superconductor electrodes separated by a barrier and coupled by a weak link. 
The former is a macroscopic quantum phenomenon where a current could flow for a long period of time without any voltage supply.
It was British physicist Brian David Josephson who investigated the relationship between the current and voltage across that weak link~\cite{Josephson62,Josephson74}.
Reviews on Josephson junctions and superconducting soliton oscillators in the context of the sine-Gordon model are given by~\cite{Parmentier93,Pnevmatikos93}.

The derivation of the NLSE from the sine-Gordon system for a small-amplitude limit was considered by~\cite{Kaup78} who discovered a phase-locked breather with an applied alternating current field by means of perturbation theory. Nonlinear breather dynamics of an alternating current parametric force in the presence of loss in a sine-Gordon system has been analyzed by~\cite{Gronbech93}. In the case of a small-amplitude limit where the system can be described by an effective NLSE, a correct threshold value for the driving force amplitude was obtained when the breather frequency was identically one. 

Inspired by the recent progress in quantum graph theory and its applications~\cite{Gnutzmann06,Kuchment08,Berkolaiko13}, interactions of traveling localized wave solutions with a vertex in a star graph from a tricrystal Josephson junction has been investigated recently by~\cite{Susanto19}. Other applications of the sine-Gordon and nonlinear Klein-Gordon models include a mechanical model with springs, wires and bearings, and Bloch wall dynamics in magnetic crystals~\cite{Barone71}. For a summary in contemporary developments of the sine-Gordon model and its wide range of applications, please consult~\cite{Cuevas14}.

\section{Nonlinear optics}   \label{nonlinearoptics}

\subsection{NLSE derivation from Maxwell's and Helmholtz' equations}

The derivation in this subsection follows the argument presented by~\cite{Agrawal12,Kivshar03,Sulem99}. See also~\cite{Moloney19,Banerjee04,Butcher90}. 
Maxwell's equations govern the propagation of electromagnetic waves and optical fields in fibers, given as follows in the International System of Units:
\begin{align}
\nabla \times \mathbf{E} &= -\partial_{t}\mathbf{B}             \qquad \qquad \quad \text{(Faraday's law)} \label{faraday} \\
\nabla \times \mathbf{H} &= \mathbf{J} + \partial_{t}\mathbf{D} \qquad \qquad       \text{(Ampere's law)}  \label{ampere} \\
\nabla \cdot  \mathbf{D} &= \rho \\
\nabla \cdot  \mathbf{B} &= 0.
\end{align}
Here, $\mathbf{E}$ and $\mathbf{H}$ denote electric and magnetic vector fields, respectively; 
$\mathbf{D}$ and $\mathbf{B}$ denote the corresponding electric and magnetic flux densities;
$\rho$ is the charge density and $\mathbf{J}$ is the corresponding current density of free charges and both represent the sources for the electromagnetic field.
We also have the following constitutive relations for electric and magnetic flux densities:
\begin{align}
\mathbf{D} &= \epsilon_0 \mathbf{E} + \mathbf{P} 		  \label{dep} \\
\mathbf{B} &= \mu_0 \mathbf{H} + \mathbf{M}               \label{bhm}
\end{align}
where $\epsilon_0$ is the vacuum permittivity, $\mu_0$ is the vacuum permeability, and $\mathbf{P}$ and $\mathbf{M}$ are the induced electric and magnetic polarizations.
We will adopt the following common assumptions in nonlinear fiber optics: the absence of free charges ($\rho = 0$ and $\mathbf{J} = \mathbf{0}$) and a nonmagnetic medium like fiber optics ($\mathbf{M} = \mathbf{0}$). 
By taking the curl of Faraday's law~\eqref{faraday}, using~\eqref{bhm},~\eqref{ampere}, and~\eqref{dep}, we can eliminate $\mathbf{B}$ and $\mathbf{D}$ to obtain an expression in $\mathbf{E}$ and $\mathbf{P}$
\begin{equation}
\nabla \times \nabla \times \mathbf{E} + \frac{1}{c^2} \partial_{t}^2 \mathbf{E} = -\mu_0 \partial_{t}^2 \mathbf{P}, \label{maxwell}
\end{equation} 
where $1/c^2 = \mu_0 \epsilon_0$ is the speed of light in a vacuum.

We adopt a common relationship between the induced polarization $\mathbf{P}$ and the electric field $\mathbf{E}$ which is valid in the electric-dipole approximation and under the assumption of local medium response. The induced polarization $\mathbf{P}$ is written as the combination of the linear and the nonlinear parts, $\mathbf{P}_L$ and $\mathbf{P}_{NL}$, respectively: $\mathbf{P}(\mathbf{r},t) = \mathbf{P}_L(\mathbf{r},t) + \mathbf{P}_{NL}(\mathbf{r},t)$, where
\begin{align}
\mathbf{P}_L   (\mathbf{r},t) &= \epsilon_0 \int_{-\infty}^{t} \chi^{(1)} (t - t_0) \mathbf{E}(\mathbf{r},t_0) \, \mathrm{d}t_0 \\
\mathbf{P}_{NL}(\mathbf{r},t) &= \epsilon_0 \int_{-\infty}^{t} \int_{-\infty}^{t} \int_{-\infty}^{t} \chi^{(3)} (t - t_1,t - t_2, t - t_3) \mathbf{E}(\mathbf{r},t_1) \mathbf{E}(\mathbf{r},t_2) \mathbf{E}(\mathbf{r},t_3) \, \mathrm{d}t_1 \, \mathrm{d}t_2 \, \mathrm{d}t_3
\end{align}
where $\chi^{(j)}$ is a tensor of rank $j + 1$, the $j$-th order of susceptibility.
Consider the case where $\mathbf{P}_{NL} = \mathbf{0}$. Let $\mathbf{\hat{E}}(\mathbf{r},\omega)$ be the Fourier transform of $\mathbf{E}(\mathbf{r},t)$, defined as
\begin{equation}
\mathbf{\hat{E}}(\mathbf{r},\omega) = \int_{-\infty}^{\infty} \mathbf{E}(\mathbf{r},t) \, e^{i\omega t} \, \mathrm{d}t
\end{equation}
and let $\hat{\chi}^{(1)}(\omega)$ be the Fourier transform of the linear, first-order susceptibility of $\chi^{(1)}(t)$, then~\eqref{maxwell} can be written in the frequency domain
\begin{equation}
\nabla \times \nabla \times \hat{\mathbf{E}} = \epsilon(\omega) \frac{\omega^2}{c^2} \mathbf{\hat{E}}(\mathbf{r},\omega)
\end{equation}
where $\epsilon(\omega) = 1 + \hat{\chi}^{(1)}(\omega) = \left(n + i \alpha c/(2 \omega)\right)^2$ is the frequency-dependent dielectric constant and its real and imaginary parts are related to the refractive index $n(\omega)$ and the absorption coefficient $\alpha(\omega)$. 

Due to low optical loses in fibers within the wavelength region of interest, Im$\{\chi^{(1)}(\omega)\} \ll \text{Im}\{\chi^{(1)}(\omega)\}$ and hence, $\epsilon(\omega) \approx n^2(\omega)$. In addition, since usually $n(\omega)$ is independent of spatial coordinates, then $\nabla \cdot \mathbf{D} = \epsilon \, \nabla \cdot \mathbf{E} = 0$ and hence, $\nabla \times \nabla \times \mathbf{E} = \nabla (\nabla \cdot \mathbf{E}) - \nabla^2 \mathbf{E} = - \nabla^2 \mathbf{E}$. Finally, we arrive at the Helmholtz equation
\begin{equation}
\nabla^2 \mathbf{\hat{E}} + n^2 (\omega) \frac{\omega^2}{c^2} \mathbf{\hat{E}} = 0.  \label{helmholtz1}
\end{equation}
By including the nonlinear effect of the induced polarization, the Helmholtz equation can be written as
\begin{equation}
\nabla^2 \mathbf{\hat{E}} + \epsilon (\omega) k_0^2 \mathbf{\hat{E}} = 0             \label{helmholtz2}
\end{equation}
where $k_0 = \omega/c$ and the dielectric constant $\epsilon(\omega) = 1 + \hat{\chi}^{(1)}(\omega) + \frac{3}{4} \frac{d^4 \chi^{(3)}}{dx^4} |E(\mathbf{r},t)|^2$.
The Helmholtz equation~\eqref{helmholtz2} can be solved using the method of separation of variables.
We assume an Ansatz in the following form:
\begin{equation}
\hat{E}(\mathbf{r},\omega - \omega_0) = \hat{A}(z, \omega - \omega_0) \, B(x,y) \, e^{i \beta_0 z}
\end{equation}
where $\hat{A}(z,\omega)$ is a slowly varying function of $z$ and $\beta_0$ is a wavenumber that needs to be determined.
The Helmholtz equation~\eqref{helmholtz2} leads to the following equations for $\hat{A}(z,\omega)$ and for $B(x,y)$, where we have neglected $\partial_{z}^2 \hat{A}$ due to an assumption of slowly varying function in $z$:
\begin{align}
2i \beta_0 \partial_{z} \hat{A} + (\hat{\beta}^2 - \beta_0^2)^2 \hat{A} &= 0     \label{eigenafou}\\
\nabla^2 B + \left[\epsilon(\omega) k_0^2 - \hat{\beta}^2 \right] B &= 0.        \label{eigenbeta}
\end{align}
The wavenumber $\hat{\beta}$ is determined by solving the eigenvalue equation~\eqref{eigenbeta} using the first-order perturbation theory.
We obtain
\begin{equation}
\hat{\beta}(\omega) = \beta(\omega) + \Delta \beta 					\label{betaeigen}
\end{equation}
where
\begin{equation}
\Delta \beta = \frac{\omega^2 n(\omega)}{c^2 \beta(\omega)} \frac{\displaystyle \int_{\infty}^{\infty} \int_{-\infty}^{\infty} \Delta n(\omega) |B(x,y)|^2 \, \mathrm{d}x \, \mathrm{d}y}{\displaystyle \int_{\infty}^{\infty} \int_{-\infty}^{\infty} |B(x,y)|^2 \, \mathrm{d}x \, \mathrm{d}y}.
\end{equation}
Using~\eqref{betaeigen} and approximating $\hat{\beta}^2 - \beta_0^2 \approx 2 \beta_0 (\hat{\beta} - \beta_0)$, the Fourier transform $\hat{A}(z,\omega - \omega_0)$ satisfying~\eqref{eigenafou} can be written as follows:
\begin{equation}
\partial_{z} \hat{A} = i \left[ \beta(\omega) + \Delta \beta(\omega) - \beta_0 \right] \hat{A}.
\end{equation}
Since an exact form of the propagation constant $\beta(\omega)$ is rarely known, it is beneficial to expand both $\beta(\omega)$ and $\Delta \beta(\omega)$ in a Taylor series about the carrier frequency $\omega_0$
\begin{align}
\beta(\omega) &= \beta(\omega_0) + \beta'(\omega_0) (\omega - \omega_0) + \frac{1}{2} \beta''(\omega_0) (\omega - \omega_0)^2 + \dots \\
\Delta \beta(\omega) &= \Delta \beta(\omega_0) + \Delta \beta'(\omega_0) (\omega - \omega_0) + \frac{1}{2} \Delta \beta''(\omega_0) (\omega - \omega_0)^2 + \dots.
\end{align}
Replacing $\omega - \omega_0$ with the differential operator $i\partial_{t}$ and taking back the inverse Fourier transform of $\hat{A}(z,\omega - \omega_0)$, we obtain the following equation for $A(z,t)$:
\begin{equation}
i \partial_{z} A + i \beta'(\omega_0) \partial_{t} A - \frac{1}{2} \beta''(\omega_0) \partial_{t}^2 A + \Delta \beta(\omega_0) A = 0.
\end{equation}
Using the transformation of a moving frame of reference $T = t - \beta'(\omega_0) z$ and considering that the last term contains the fiber loss and nonlinearity effects, we obtain the NLSE
\begin{equation}
i \partial_{z} A - \frac{1}{2} \beta''(\omega_0) \partial_{T}^2 A + \gamma |A|^2 A = 0.  \label{NLSEtemporalopticalsoliton}
\end{equation}
Here, the nonlinear coefficient $\gamma$ is given by
\begin{equation}
\gamma(\omega_0) = -\frac{\omega_0}{c} n_2(\omega_0) \frac{\displaystyle \int_{\infty}^{\infty} \int_{-\infty}^{\infty} |B(x,y)|^4 \, \mathrm{d}x \, \mathrm{d}y}{\displaystyle \left(\int_{\infty}^{\infty} \int_{-\infty}^{\infty} |B(x,y)|^2 \, \mathrm{d}x \, \mathrm{d}y \right)^2}.
\end{equation}
For a single-mode fiber, the modal distribution $B(x,y)$ corresponds to the fundamental fiber mode, given by one of the following expressions:
\begin{align}
B(x,y) &= \left\{
\begin{array}{ll}
J_0(p \sqrt{x^2 + y^2}),  & \qquad \sqrt{x^2 + y^2} \leq a \\
\frac{\sqrt{a}}{\sqrt[4]{x^2 + y^2}} J_0(pa) e^{-q(\sqrt{x^2 + y^2} - a)}, & \qquad \sqrt{x^2 + y^2} \geq a
\end{array}
\right. \\
\text{or} \quad B(x,y) &= e^{-\frac{(x^2 + y^2)}{w^2}}.
\end{align}
Here, $J_0$ denotes the Bessel function of the first kind of order zero, $a$ is the radius of the fiber core, $w$ is a width parameter, and the quantities $p = \sqrt{n_1^2 k_0^2 - \beta^2}$ and $q = \sqrt{\beta^2 - n_c^2 k_0^2}$.

\subsection{Applications in nonlinear optics}

In the context of nonlinear optics, an of interest object of study is ``solitary waves,'' also known as ``solitons.''
Depending on whether the light confinement occurs in time or space during wave propagation, solitons can be classified as either temporal or spatial.
The NLSE~\eqref{NLSEtemporalopticalsoliton} governs the time-dependent pulse envelope propagation in optical fibers, also known as temporal soliton.
On the other hand, the spatial soliton, a continuous wave beam propagation inside a nonlinear optical medium with Kerr (or cubic) nonlinearity, is governed by the following $(1 + 1)$D NLSE
\begin{equation}
i\partial_{z}A + \beta \partial_{X}^2 A + \gamma |A|^2 A = 0. \label{NLSEspatialopticalsoliton}
\end{equation}
A classical example for the latter is a bell-shaped spatial wavepacket with the self-induced lensing effect, a phenomenon of self-trapping in dielectric waveguide modes discovered by~\cite{Chiao64}. A stable spatial soliton was also observed experimentally with self-trapping laser beams propagating through homogeneous transparent dielectrics~\cite{Barthelemy85}. For an extensive coverage on spatial solitons, please consult~\cite{Trillo01}.

In the following, we will discuss temporal solitons in optical fibers.
An optical fiber is a flexible and transparent material fiber made by drawing a pure silica glass or plastic fiber. 
The center core is surrounded by an outer layer, known as cladding, with lower refractive index optical material than the core.
The fiber is then coated with buffer and jacket for protection from moisture and physical damage. 
Optical fibers are widely used in fiber-optics communications as a means to transmit light between the two ends of the fiber.
The light remains in the core due to total internal reflection so that the fiber acts as a waveguide.

The result of an interaction between fiber dispersion and nonliearity is fiber optic solitons.
The self-phase modulation phenomenon, where an ultrashort pulse of light inducing a varying refractive index when traveling in a medium due to the optical Kerr effect, can balance the anomalous group velocity dispersion with a nonlinearity effect to create an optical soliton.
An existence of temporal optical solitons in the context of optical fibers was predicted theoretically by~\cite{Hasegawa73a,Hasegawa73b} and experimentally confirmed by~\cite{Molleanauer80}.
These solitons represent optical pulses that maintain their shape during propagation and belong to families of exact solution of the NLSE.

The field of nonlinear fiber optics continues to progress with the development of Erbium-doped fiber amplifiers~\cite{Becker99,Desurvire02}.
With the advent of this century, new types of fiber optic amplifiers indicating nonlinear effects were developed, including stimulated Raman scattering and four-wave mixing~\cite{Headley05,Pal10}.
This further leads to other types of solitons such as dispersion-managed solitons and dissipative solitons~\cite{Hasegawa03,Kivshar03,Akhmediev05,Mollenauer06,Akhmediev08}.
For feature articles on theoretical and experimental challenges in optical solitons, the readers may consult a volume edited by~\cite{Porsezian03}.

In addition to the well-known bright, dark and gray solitons, ``optical rogue waves'' have gained popularity during the past decade in the field of nonlinear optics.
The experimental result on rogue waves in an optical system is supported by numerical simulation based on probabilistic supercontinuum generation in a highly nonlinear microstructured optical fiber and a generalized NLSE as a mathematical model~\cite{Solli07,Dudley08,Bonatto11}. An overview on the research dynamics in optical rogue waves and the state of the art on the subject has been covered by~\cite{Akhmediev13}. For discussion and debate on whether the science of rogue waves is moving towards a unifying concept, please consult the papers published by various authors in \textit{The European Physical Journal Special Topics}, Volume 185, pages 1--266, July 2010, published by Springer-Verlag.

\section{Bose-Einstein condensation} \label{BEcondensate}

\subsection{NLSE derivation from Bose-Einstein condensed state}

A Bose-Einstein condensate (BEC) is a state of matter of a low density dilute gas, also known as bosons, for which cooling down to a nearly absolute zero temperature would cause them to condense into the lowest accessible quantum state. The phenomenon was predicted nearly a century ago by~\cite{Bose24} and~\cite{Einstein24}.
In the context of BEC, the NLSE is known as Gross-Pitaevskii equation (GPE), where the model was derived independently by~\cite{Gross61} and~\cite{Pitaevskii61}. 
The derivation presented in this section follows an argument presented in~\cite{Pitaevskii03,Pethick08,Dalfovo99}.
For a rigorous treatment of the modeling, see~\cite{Erdos07,Erdos10}. 

Consider a system of a weakly-interacting Bose gas where its Hamiltonian can be written in terms of the field operator $\psi$
\begin{equation}
\hat{H} = \int \left(\frac{\hbar}{2m} \nabla \hat{\Psi}^{\dagger} \nabla \hat{\Psi} \right) \mathrm{d}\mathbf{r} + \frac{1}{2} \int \hat{\Psi}^{\dagger} \hat{\Psi}^{\dagger \prime} V(\mathbf{r}' - \mathbf{r}) \hat{\Psi} \hat{\Psi}^{\prime} \, \mathrm{d}\mathbf{r}^{\prime} \, \mathrm{d}\mathbf{r}.  \label{hamil}
\end{equation}
Here, $\hbar$ is the Planck constant, $m$ is particle mass, $\hat{\Psi}^{\dagger}(\mathbf{r})$ and $\hat{\Psi}(\mathbf{r})$ are the field operator creating and annihilating a particle at the point $\mathbf{r}$, and $V(\mathbf{r})$ is the two-body potential.
The field operators satisfy the following commutation relationship
\begin{equation}
\left[\hat{\Psi}(\mathbf{r}), \hat{\Psi}^{\dagger}(\mathbf{r}) \right] = \delta(\mathbf{r} - \mathbf{r}') \quad \quad \text{and} \quad \quad
\left[\hat{\Psi}(\mathbf{r}), \hat{\Psi}(\mathbf{r}^\prime) \right] = 0. \label{commu}
\end{equation}
In the Heisenberg representation, the field operator $\hat{\Psi}(\mathbf{r},t)$ satisfies
\begin{align}
i \hbar \partial_t \hat{\Psi}(\mathbf{r},t) &= \left[\hat{\Psi}(\mathbf{r},t), \hat{H} \right] \\
&= \left[- \frac{\hbar^2 \nabla^2}{2 m} + V_\text{ext}(\mathbf{r},t) + \int \hat{\Psi}^{\dagger}(\mathbf{r}^\prime, t) V(\mathbf{r}^\prime - \mathbf{r}) \hat{\Psi}(\mathbf{r}^\prime,t) \, \mathrm{d}\mathbf{r}^\prime \right] \, \hat{\Psi}(\mathbf{r},t)
\end{align}
where $V_\text{ext}$ is an external potential and we have utilized the Hamiltonian~\eqref{hamil} and the commutation relationship~\eqref{commu}.
If we apply an effective potential $V_\text{eff}$ where the Born approximation is applicable~\cite{Born26}, then we can replace the field operator $\hat{\Psi}(\mathbf{r},t)$ with a classical field or the condensation wave function $\Psi_0(\mathbf{r},t)$ at a very low temperature and up to the lowest-order approximation.
Under an assumption of the slowly varying function $\Psi_0(\mathbf{r},t)$ on distances of the order of interatomic force range, $\mathbf{r}^\prime$ can be replaced with $\mathbf{r}$.
We arrive at the GPE
\begin{equation}
i \hbar \partial_t \Psi_0(\mathbf{r},t) = \left(-\frac{\hbar^2 \nabla^2}{2 m} + V_\text{ext}(\mathbf{r},t) + g|\Psi_0(\mathbf{r},t)|^2 \right) \Psi_0(\mathbf{r},t),
\end{equation}
where ${\displaystyle g = \int V_\text{eff}(\mathbf{r}) \, \mathrm{d}\mathbf{r}}$.
This GPE governs the ground state of a quantum system of identical bosons where it is used as a model equation for the single-particle wavefunction in a BEC.
In particular, the presence of external potential $V_\text{ext}$ allows us to model various situations of the external world action on the condensate. 

\subsection{Applications in BEC}

Although the low-temperature and high-density state of BEC was predicted in the 1920s, it was not until the 1990s that the phenomenon has been successfully implemented and experimentally tested in laboratories, by confining in magnetic traps and cooling down to extremely low temperatures, vapors of rubidium $^{87}$Rb~\cite{Anderson95}, lithium $^{7}$Li~\cite{Bradley95}, and sodium $^{23}$Na~\cite{Davis95} atoms. 
An intensive effort on other atomic species has also produced fruitful experimental results, such as on dilute gasses of atomic hydrogen~\cite{Fried98,Greytak00} and helium in the $2^3 \, S_1$ metastable state $^4$He$^{\ast}$~\cite{Dos01,Robert01}, as well as samples of potassium $^{41}$K~\cite{Modugno01}, cesium $^{133}$Cs~\cite{Weber03}, and another isotope of rubidium $^{85}$Rb~\cite{Cornish00}.
The group at the Joint Institute for Laboratory Astrophysics in Colorado successfully measured, for the first time, the collective excitations of a Bose condensed dilute gas in a trap~\cite{Jin96}.

For the modeling aspect, the dynamics of a dilute trapped BEC has been successfully constructed using the mean-field theory, within the self-consistent~\cite{Hartree28}-\cite{Fock30}-\cite{Bogo47} approximation, where indeed the GPE can also be derived~\cite{Griffin96}. 
The majority of theoretical approach in solving the GPE is centered around the Thomas-Fermi approximation~\cite{Thomas27,Fermi27}, where the nonlinear atomic and interactions are much larger than the kinetic energy pressure, and hence the latter is neglected~\cite{Baym96,Stringari96,Dalfovo96}. 
An analytical attempt in solving the GPE to model the dynamics of dilute ultracold atom clouds in the BEC phase by applying a variational technique is proposed by~\cite{Perez96,Perez97}. 

The GPE for BEC has also been studied and solved numerically, by direct numerical integration for time-independent~\cite{Edwards95} and time-dependent GPEs~\cite{Ruprecht96}. Other techniques for the latter include, but are not limited to, the semi-implicit Crank-Nicholson scheme~\cite{Ruprecht95}, an eigenfunction basis expansion method~\cite{Edwards96}, an explicit finite-difference scheme~\cite{Cerimele00}, and the time-splitting spectral method~\cite{Bao03}.

Another application of the GPE to a 1D cloud boson is covered by~\cite{Roger13}. This 1D bosonic gas with a uniform potential can be obtained by getting a condensate in an elongated trap.
In particular, the phenomena occurring at the center of the trap where the density is relatively uniform and the condensate behaves as a fluid is interesting. For repulsive interactions, the solution of the GPE is in the form of a hyperbolic tangent profile while for attractive interactions, it becomes a hyperbolic secant profile.

An extensive review of the mean-field theory applied to BEC is given by~\cite{Dalfovo99}. See also other reviews by~\cite{Burnett96} and~\cite{Parkins98}.
For an excellent symbiosis between theoretical and experimental contributions on BEC, the readers are encouraged to consult~\cite{Kevrekidis08}.

\section{Conclusion}  \label{conclusion}

In this chapter, we have provided an overview of modeling and application aspects of the NLSE in various physical settings. Indeed, the subject of NLSE is an active and dynamic research area not only in Mathematical Physics, but also in other areas of Science as well as in Engineering. We have derived the NLSE heuristically as well as by implementing the method of multiple-scales. We also covered the applications of NLSE in surface water waves, superconductivity, nonlinear optics and BEC, including solitons and rogue waves. The evolution equation admittedly has some limitations, yet, it is remarkable that the NLSE provides a rather universal model in several areas that do not seem to be closely connected according to a general point of view. We hope that this chapter will stimulate further research on these exciting topics.

\addcontentsline{toc}{section}{Acknowledgement}
\section*{\large Acknowledgements}
{\small The author would like to acknowledge Professor E. (Brenny) W. C. van Groesen (University of Twente, the Netherlands and LabMath Indonesia), Professor Hadi Susanto (The University of Essex, UK), Professor Nail N. Akhmediev (Australian National University), Professor Panayotis G. Kevrekidis (University of Amherst, Massachusetts), Professor Mason A. Porter (University of California, Los Angeles), Professor Guido Schneider (University of Stuttgart, Germany), Dr. Gert Klopman (Witteveen$+$Bos, the Netherlands), Dr. Agung Trisetyarso (Bina Nusantara University, Indonesia), Drs. Andonowati, Alexander A. P. Iskandar, and Rudy Kusdiantara (Bandung Institute of Technology, Indonesia), Professor Fr\'ed\'eric Dias (University College Dublin, Ireland), Professor Kristian B. Dysthe (University of Bergen, Norway), Professor Karsten Trulsen (University of Oslo, Norway), Professors Miguel Onorato and Alfred R. Osborne (Universit\'a di Torino, Italy), Dr. Ardhasena Sopaheluwakan (Meteorology, Climatology, and Geophysical Agency, Indonesia), Professor Christian Kharif (\textit{Institut de Recherche sur les Ph\'enom\`enes Hors Equilibre}, Research Institute for Non-Equilibrium Phenomena, Aix- Marseille University, France), Professors Ren\'e H. M. Huijsmans and Jurjen A. Battjes (Delft University of Technology, the Netherlands), Professor Arthur E. Mynett (International Institute for Hydraulic and Environmental Engineering IHE Delft Institute for Water Education and Deltares, formerly WL|Delft Hydraulics, the Netherlands), and Dr. Pearu Peterson (Tallinn University of Technology, Estonia) for advices, suggestions, and fruitful discussions. This research is supported by the Dutch Organization of Scientific Research NWO (\textit{Nederlandse Organisatie voor Wetenschappelijk Onderzoek}), subdivision Applied Sciences STW (\textit{Stichting Technische Wetenschappen}) through Mathematics and Innovation Transfer Point (\textit{Transferpunt Wiskunde en Innovatie}) Grant No. TWI-5374, the New Researcher Fund from the University of Nottingham,  University  Park  Campus,  UK, through Grant No. NRF 5035-A2RL20, the SKKU Samsung Intramural Research Fund No. 2016-1299-000, and the Beginning Independent Researcher Program ({\slshape Saengae-Cheot Yeongu}) from the National Research Foundation of Korea through Grant No. NRF-2017-R1C1B5-017743 under the Basic Research Program in Science and Engineering.\par}


\label{lastpage}

\begin{thebibliography}{99}
\bibitem[Ablowitz et al 1974]{Ablowitz74} Ablowitz, Mark J., David James Kaup, Alan C. Newell, and Harvey Segur. 1974. ``The inverse scattering transform-Fourier analysis for nonlinear problems''. \textit{Studies in Applied Mathematics} 53, no. 4: 249--315.

\bibitem[Ablowitz and Segur 1981]{Ablowitz81} Ablowitz, Mark J., and Harvey Segur. 1981. \textit{Solitons and the Inverse Scattering Transform}. Vol. 4. Philadelphia, PA: Society for Industrial and Applied Mathematics.

\bibitem[Ablowitz and Herbst 1990]{Ablowitz90} Ablowitz, Mark J., and B. M. Herbst. 1990. ``On homoclinic structure and numerically induced chaos for the nonlinear Schr\"odinger equation''. \textit{SIAM Journal on Applied Mathematics} 50, no. 2: 339--351.
\addcontentsline{toc}{section}{References}

\bibitem[Ablowitz and Clarkson 1991]{Ablowitz91} Ablowitz, Mark J., and Peter A. Clarkson. 1991. \textit{Solitons, Nonlinear Evolution Equations and Inverse Scattering}. Vol. 149. Cambridge, UK: Cambridge University Press.

\bibitem[Agrawal 2012]{Agrawal12} Agrawal, Govind P. 2012. \textit{Nonlinear Fiber Optics, Fifth Edition}. Burlington, MA: Academic Press.

\bibitem[Akhmediev, Eleonskii and Kulagin 1987]{Akhmediev87} Akhmediev, Nail N., Vladimir Markovich Eleonskii, and N. E. Kulagin. 1987. ``Exact first-order solutions of the nonlinear Schr\"{o}dinger equation''. \textit{Theoretical and Mathematical Physics} 72, no. 2: 809--818.

\bibitem[Akhmediev and Ankiewicz 1997]{Akhmediev97} Akhmediev, Nail N., and Adrian Ankiewicz. 1997. \textit{Solitons: Nonlinear Pulses and Beams}. London, UK: Chapman \& Hall. 

\bibitem[Akhmediev and Ankiewicz 2005]{Akhmediev05} Akhmediev, Nail N., and Adrian Ankiewicz. 2005. \textit{Dissipative Solitons}. Berlin Heidelberg, Germany: Springer-Verlag.

\bibitem[Akhmediev and Ankiewicz 2008]{Akhmediev08} Akhmediev, Nail N., and Adrian Ankiewicz. 2008. \textit{Dissipative Solitons: From Optics to Biology and Medicine}. Berlin Heidelberg, Germany: Springer-Verlag.

\bibitem[Akhmediev et al 2013]{Akhmediev13} Akhmediev, Nail, John M. Dudley, Daniel R. Solli, and S. K. Turitsyn. 2013. ``Recent progress in investigating optical rogue waves''. \textit{Journal of Optics} 15, no. 6: 060201.

\bibitem[Anderson et al 1995]{Anderson95} Anderson, Mike H., Jason R. Ensher, Michael R. Matthews, Carl E. Wieman, and Eric A. Cornell. 1995. ``Observation of Bose-Einstein condensation in a dilute atomic vapor''. \textit{Science} 269: 198--201.

\bibitem[Andonowati, Karjanto and van Groesen 2007]{Andonowati07} Andonowati, Natanael Karjanto, and Embrecht W. C. van Groesen. 2007. ``Extreme wave phenomena in down-stream running modulated waves''. \textit{Applied Mathematical Modelling} 31, no. 7: 1425--1443.

\bibitem[Banerjee 2004]{Banerjee04} Banerjee, Partha P. 2004. \textit{Nonlinear Optics--Theory, Numerical Modeling, and Applications}. New York, NY: Marcel Dekker.

\bibitem[Bao, Jaksch and Markowich 2003]{Bao03} Bao, Weizhu, Dieter Jaksch, and Peter A. Markowich. 2003. ``Numerical solution of the Gross--Pitaevskii equation for Bose--Einstein condensation''. \textit{Journal of Computational Physics} 187, no. 1: 318--342.

\bibitem[Barone et al 1971]{Barone71} Barone, A., F. Esposito, C. J. Magee, and A. C. Scott. 1971. ``Theory and applications of the sine-Gordon equation''.  \textit{La Rivista del Nuovo Cimento (1971-1977) (The Journal of New Cimento)} 1, no. 2: 227--267.

\bibitem[Barthelemy, Maneuf and Froehly 1985]{Barthelemy85} Barthelemy, Alain, Serge Maneuf, and Claude Froehly. 1985. ``Propagation soliton et auto-confinement de faisceaux laser par non linearit\'e optique de Kerr (Soliton propagation and self-trapping of laser beams by a Kerr optical nonlinearity)''. \textit{Optics Communications} 55, no. 3: 201--206.

\bibitem[Baym and Pethick 1996]{Baym96} Baym, Gordon, and C. J. Pethick. 1996. ``Ground-state properties of magnetically trapped Bose-condensed rubidium gas''. \textit{Physical Review Letters} 76, no. 1: 6--9.

\bibitem[Becker, Olsson and Simpson 1999]{Becker99} Becker, Philippe C., N. Anders Olsson, and Jay R. Simpson. 1999. \textit{Erbium-Doped Fiber Amplifiers: Fundamentals and Technology}. San Diego, CA: Academic Press and Amsterdam, the Netherlands: Elsevier.

\bibitem[Benjamin and Feir 1967]{Benjamin67} Benjamin, Thomas Brooke, and James E. Feir. 1967. ``The disintegration of wave trains on deep water Part 1. Theory.'' \textit{Journal of Fluid Mechanics} 27, no. 3: 417--430.

\bibitem[Benjamin 1967]{Benjamin67a} Benjamin, Thomas Brooke. 1967. ``Instability of periodic wavetrains in nonlinear dispersive systems''. \textit{Proceedings of the Royal Society of London. Series A. Mathematical and Physical Sciences} 299, no. 1456: 59--76.

\bibitem[Berkolaiko and Kuchment 2013]{Berkolaiko13} Berkolaiko, Gregory, and Peter Kuchment. 2013. \textit{Introduction to Quantum Graphs}. Vol. 186 of Mathematical Surveys and Monographs. Providence, RI: American Mathematical Society.

\bibitem[Bludov et al 2009]{Bludov09} Bludov, Yu V., V. V. Konotop, and Nail Akhmediev. 2009. ``Matter rogue waves''. \textit{Physical Review A} 80, no. 3: 033610.

\bibitem[Bogoliubov 1947]{Bogo47} Bogoliubov, Nikolay Nikolayevich. 1947. ``On the theory of superfluidity''. \textit{Journal of Physics (Moscow)} 11, no. 1: 23--32.

\bibitem[Bonatto et al 2011]{Bonatto11} Bonatto, Cristian, Michael Feyereisen, St\'ephane Barland, Massimo Giudici, Cristina Masoller, Jos\'e R. Rios Leite, and Jorge R. Tredicce. 2011. ``Deterministic optical rogue waves''. \textit{Physical Review Letters} 107, no. 5: 053901.

\bibitem[Born 1926]{Born26} Born, Max. 1926. ``Quantenmechanik der Sto{\ss}vorg\"ange''. (``Quantum mechanics of the collision processes''). \textit{Zeitschrift f\"ur Physik (Journal of Physics)} 38, no. 11-12: 803--827.

\bibitem[Bose 1924]{Bose24} Bose, Satyendra Nath. 1924. ``Plancks gesetz und lichtquantenhypothese''. (``Planck's law and light quantum hypothesis''). \textit{Zeitschrift f\"ur Physik (Journal of Physics)} 26, no. 1: 178--181.

\bibitem[Bourgain 1999]{Bourgain99} Bourgain, Jean. 1999. \textit{Global Solutions of Nonlinear Schr\"odinger Equations}. Vol. 46. Providence, RI: American Mathematical Society.

\bibitem[Bradley et al 1995]{Bradley95} Bradley, C. C., C. A. Sackett, J. J. Tollett, and Randall G. Hulet. 1995. ``Evidence of Bose-Einstein condensation in an atomic gas with attractive interactions''. \textit{Physical Review Letters} 75, no. 9: 1687--1690.

\bibitem[Burnett 1996]{Burnett96} Burnett, Keith. 1996. ``Bose-Einstein condensation with evaporatively cooled atoms''. \textit{Contemporary Physics} 37, no. 1: 1--14.

\bibitem[Butcher and Cotter 1990]{Butcher90} Butcher, Paul N., and David Cotter. 1990. \textit{The Elements of Nonlinear Optics}. Vol. 9. Cambridge, UK: Cambridge University Press.

\bibitem[Cahyono 2002]{Cahyono02} Cahyono, Edi. 2002. ``Analytical Wave Codes for Predicting Surface Waves in a Laboratory Basin''. PhD diss., University of Twente, the Netherlands. 

\bibitem[Cazenave 1996a]{Cazenave96a} Cazenave, Thierry. 1996. \textit{An Introduction to Nonlinear Schr\"{o}dinger Equations, Third Edition}. Vol. 26. \textit{Textos de M\'etodos Matem\'aticos (Mathematical Method Textbooks)}. Rio de Janeiro, Brazil: Universidade Federal do Rio de Janeiro, Centro de Ci\^{e}ncias Matem\'aticas e da Natureza, Instituto de Matem\'atica (Federal University of Rio de Janeiro, Center for Mathematical and Natural Science, Institute of Mathematics).

\bibitem[Cazenave 1996b]{Cazenave96b} Cazenave, Thierry. 1996. \textit{Blow up and Scattering in the Nonlinear Schr\"odinger Equation, Second Edition}. Vol. 30. \textit{Textos de M\'etodos Matem\'aticos (Mathematical Method Textbooks)}. Rio de Janeiro, Brazil: Universidade Federal do Rio de Janeiro, Centro de Ci\^{e}ncias Matem\'aticas e da Natureza, Instituto de Matem\'atica (Federal University of Rio de Janeiro, Center for Mathematical and Natural Science, Institute of Mathematics).

\bibitem[Cerimele et al 2000]{Cerimele00} Cerimele, Maria Mercede, Maria Luisa Chiofalo, Francesca Pistella, Sauro Succi, and Mario P. Tosi. 2000. ``Numerical solution of the Gross-Pitaevskii equation using an explicit finite-difference scheme: An application to trapped Bose-Einstein condensates''. \textit{Physical Review E} 62, no. 1: 1382--1389.

\bibitem[Chabchoub et al 2011]{Chabchoub11} Chabchoub, A., N. P. Hoffmann, and Nail Akhmediev. 2011. ``Rogue wave observation in a water wave tank''. \textit{Physical Review Letters} 106, no. 20: 204502.

\bibitem[Chiao, Garmire and Townes 1964]{Chiao64} Chiao, Raymond Y., E. Garmire, and Charles H. Townes. 1964. ``Self-trapping of optical beams''. \textit{Physical Review Letters} 13, no. 15: 479--482.

\bibitem[Chu and Mei 1970]{Chu70} Chu, Vincent H., and Chiang C. Mei. 1970. ``On slowly-varying Stokes waves''. \textit{Journal of Fluid Mechanics} 41: 873--887.

\bibitem[Chu and Mei 1971]{Chu71} Chu, Vincent H., and Chiang C. Mei. 1971. ``The non-linear evolution of Stokes waves in deep water''. \textit{Journal of Fluid Mechanics} 47: 337--351.

\bibitem[Cornish et al 2000]{Cornish00} Cornish, Simon L., Neil R. Claussen, Jacob L. Roberts, Eric A. Cornell, and Carl E. Wieman. 2000. ``Stable $^{85}$Rb Bose-Einstein condensates with widely tunable interactions''. \textit{Physical Review Letters} 85, no. 9: 1795--1798.

\bibitem[Cuevas-Maraver, Kevrekidis and Williams 2014]{Cuevas14} Cuevas-Maraver, Jes\'us, Panayotis G. Kevrekidis, and Floyd Williams. 2014. \textit{The sine-Gordon Model and Its Applications: From Pendula and Josephson Junctions to Gravity and High-Energy Physics}. Nonlinear Systems and Complexity Series. Vol. 10. Cham, Switzerland: Springer.

\bibitem[Dalfovo, Pitaevskii and Stringari 1996]{Dalfovo96} Dalfovo, Franco, Lev P. Pitaevskii, and Sandro Stringari. 1996. ``Order parameter at the boundary of a trapped Bose gas''. \textit{Physical Review A} 54, no. 5: 4213--4127.

\bibitem[Dalfovo et al 1999]{Dalfovo99} Dalfovo, Franco, Stefano Giorgini, Lev P. Pitaevskii, and Sandro Stringari. 1999. ``Theory of Bose-Einstein condensation in trapped gases''. \textit{Reviews of Modern Physics} 71, no. 3: 463--512.

\bibitem[Davis et al 1995]{Davis95} Davis, Kendall B., M.-O. Mewes, Michael R. Andrews, Nicolaas J. van Druten, Dallin S. Durfee, D. M. Kurn, and Wolfgang Ketterle. 1995. ``Bose-Einstein condensation in a gas of sodium atoms''. \textit{Physical Review Letters} 75, no. 22: 3969--3973.

\bibitem[Debnath 1994, 343--345]{Debnath94} Debnath, Lokenath. 1994. \textit{Nonlinear Water Waves}. San Diego, CA: Academic Press.

\bibitem[Debnath 2012]{Debnath12} Debnath, Lokenath. 2012. \textit{Nonlinear Partial Differential Equations for Scientists and Engineers}. New York, NY: Springer Science \& Business Media.

\bibitem[Dehghan and Shokri 2009]{Dehghan09} Dehghan, Mehdi, and Ali Shokri. 2009. ``Numerical solution of the nonlinear Klein--Gordon equation using radial basis functions''. \textit{Journal of Computational and Applied Mathematics} 230, no. 2: 400--410.

\bibitem[Desurvire et al 2002]{Desurvire02} Desurvire, Emmanuel, Dominique Bayart, Bertrand Desthieux, and S\'ebastien Bigo. 2002. \textit{Erbium-Doped Fiber Amplifiers: Device and System Developments}. Vol. 2. New York, NY: Wiley-Interscience.

\bibitem[Dingemans 1997, 888--890]{Dingemans97} Dingemans, Maarten W. 1997. \textit{Water Wave Propagation Over Uneven Bottoms}. Vol. 13. Singapore: World Scientific.

\bibitem[Dirac 1930]{Dirac30} Dirac, Paul Adrien Maurice. 1930. \textit{The Principles of Quantum Mechanics}. Oxford, UK: Oxford University Press. (The fourth edition was published in 1958 and the most recent one reprinted in 2004.)
 
\bibitem[Dodd et al 1982, 389--494]{Dodd82} Dodd, Roger K., Hedley C. Morris, J. C. Eilbeck, and J. D. Gibbon. 1982. \textit{Soliton and Nonlinear Wave Equations}. London, UK: Academic Press.

\bibitem[Dos Santos et al 2001]{Dos01} Dos Santos, F. Pereira, J\'r\'mie L\'onard, Junmin Wang, C. J. Barrelet, F. Perales, E. Rasel, C. S. Unnikrishnan, M. Leduc, and C. Cohen-Tannoudji. 2001. ``Bose-Einstein condensation of metastable helium''. \textit{Physical Review Letters} 86, no. 16: 3459--3462.

\bibitem[Drazin and Johnson 1989, 198--199]{Drazin89} Drazin, Philip G., and Robin S. Johnson. 1989. \textit{Solitons: An Introduction}. Cambridge, UK: Cambridge University Press.

\bibitem[Dudley, Genty and Eggleton 2008]{Dudley08} Dudley, John M., Go\"ery Genty, and Benjamin J. Eggleton. 2008. ``Harnessing and control of optical rogue waves in supercontinuum generation''. \textit{Optics Express} 16, no. 6: 3644--3651.

\bibitem[Duncan 1997]{Duncan97} Duncan, D. B. ``Sympletic finite difference approximations of the nonlinear Klein--Gordon equation''. 1997. \textit{SIAM Journal on Numerical Analysis} 34, no. 5: 1742--1760.


\bibitem[Dysthe and Trulsen 1999]{Dysthe99} Dysthe, Kristian B., and Karsten Trulsen. 1999. ``Note on breather type solutions of the NLS as models for freak-waves.'' \textit{Physica Scripta} T82: 48--52.

\bibitem[Dysthe, Krogstad and M\"uller 2008]{Dysthe08} Dysthe, Kristian B., Harald E. Krogstad, and Peter M\"uller. 2008. ``Oceanic rogue waves''. \textit{Annual Review of Fluid Mechanics}. 40: 287--310.

\bibitem[Edwards and Burnett 1995]{Edwards95} Edwards, Mark, and K. Burnett. 1995. ``Numerical solution of the nonlinear Schr\"odinger equation for small samples of trapped neutral atoms''. \textit{Physical Review A} 51, no. 2: 1382--1386.

\bibitem[Edwards et al 1996]{Edwards96} Edwards, Mark, R. J. Dodd, Charles W. Clark, P. A. Ruprecht, and K. Burnett. 1996. ``Properties of a Bose-Einstein condensate in an anisotropic harmonic potential''. \textit{Physical Review A} 53, no. 4: R1950--R1953.

\bibitem[Einstein 1924]{Einstein24} Einstein, Albert. 1924. ``Quantentheorie des einatomigen idealen Gases''. (``Quantum theory of ideal monoatomic gases''). 1924. \textit{Sitzungsberichte der K\"oniglich Preussischen Akademie der Wissenschaften zu Berlin, Physikalisch-Mathematische Klasse (Proceedings of the Royal Prussian Academy of Sciences at Berlin, Physics and Mathematics Section)} 1924: 261--267.

\bibitem[Enneper 1870]{Enneper70} Enneper, A. 1870. ``\"Uber asymptotische Linien'' (``About asymptotic lines''). \textit{Nachrichten von der K\"oniglichen Gesellschaft der Wissenschaften und der Georg-Augusts-Universit\"at zu G\"ottingen} (\textit{Newsletter from the Royal Society of Sciences and the Georg-August University of G\"ottingen}) 1870: 493--510.

\bibitem[Erd{\H o}s, Schlein and Yau 2007]{Erdos07} Erd{\H o}s, L{\'a}szl{\'o}, Benjamin Schlein, and Horng-Tzer Yau. 2007. ``Rigorous derivation of the Gross-Pitaevskii equation''. \textit{Physical Review Letters} 98, no. 4: 040404.

\bibitem[Erd{\H o}s, Schlein and Yau 2010]{Erdos10} Erd{\H o}s, L{\'a}szl{\'o}, Benjamin Schlein, and Horng-Tzer Yau. 2010. ``Derivation of the Gross-Pitaevskii equation for the dynamics of Bose-Einstein condensate.'' \textit{Annals of Mathematics} 172: 291--370.

\bibitem[Evans 2010]{Evans10} Evans, Lawrence C. 2010. \textit{Partial Differential Equations, Second Edition}. Providence, RI: American Mathematical Society.

\bibitem[Fermi 1927]{Fermi27} Fermi, Enrico. 1927. ``Un metodo statistico per la determinazione di alcune priorieta dell'atome''. (``A statistical method to evaluate some properties of the atom''). \textit{Rendiconti dell'Accademia Nazionale dei Lincei (Proceedings of the Lincean National Academy)} 6: 602--607.

\bibitem[Fibich 2015]{Fibich15} Fibich, Gadi. 2015. \textit{The Nonlinear Schr\"odinger Equation--Singular Solutions and Optical Collapse}. Cham, Switzerland: Springer.

\bibitem[Fock 1930]{Fock30} Fock, Vladimir. 1930. ``N\"aherungsmethode zur L\"osung des quantenmechanischen Mehrk\"orperproblems''. (``Approximation method for solving the quantum mechanical multibody problem''). \textit{Zeitschrift f\"ur Physik (Journal of Physics)} 61, no. 1-2: 126--148.

\bibitem[Fornberg and Whitham 1978]{Fornberg78} Fornberg, Bengt, and Gerald Beresford Whitham. 1978. ``A numerical and theoretical study of certain nonlinear wave phenomena''. \textit{Philosophical Transactions of the Royal Society of London. Series A, Mathematical and Physical Sciences} 289: 373--404.

\bibitem[Frenkel and Kontorova 1939]{Frenkel39} Frenkel, J., and T. Kontorova. ``On the theory of plastic deformation and twinning''. 1939. \textit{Izvestiya Akademii Nauk SSSR, Seriya Fizicheskaya (Academy of Sciences of the USSR, Journal of Physics, Moscow)} 1: 137--149.

\bibitem[Fried et al 1998]{Fried98} Fried, Dale G., Thomas C. Killian, Lorenz Willmann, David Landhuis, Stephen C. Moss, Daniel Kleppner, and Thomas J. Greytak. 1998. ``Bose-Einstein condensation of atomic hydrogen''. \textit{Physical Review Letters} 81, no. 18: 3811--3814.

\bibitem[Gibbon et al 1979]{Gibbon79} Gibbon, J. D., I. N. James, and I. M. Moroz. 1979. ``An example of soliton behaviour in a rotating baroclinic fluid''. \textit{Proceedings of the Royal Society of London. Series A, Mathematical and Physical Sciences} 367: 219--237.

\bibitem[Ginibre 1997]{Ginibre97} Ginibre, Jean. 1997. ``An introduction to nonlinear Schr\"odinger equations'', In Agemi, Rentaro, Yoshikazu Giga and Tohru Ozawa. (Eds.) \textit{Nonlinear Waves: Proceedings of the Fourth MSJ International Research Institute}, Sapporo, July 10-21, 1995, pages 80--127, GAKUTO International Series. Mathematical Sciences and Applications, Volume 10. Tokyo, Japan: Gakk\={o}tosho.

\bibitem[Gnutzmann and Smilansky 2006]{Gnutzmann06} Gnutzmann, Sven, and Uzy Smilansky. 2006. ``Quantum graphs: Applications to quantum chaos and universal spectral statistics''. \textit{Advances in Physics} 55, no. 5-6: 527--625.

\bibitem[Greytak et al 2000]{Greytak00} Greytak, T. J., D. Kleppner, D. G. Fried, T. C. Killian, L. Willmann, D. Landhuis, and S. C. Moss. 2000. ``Bose--Einstein condensation in atomic hydrogen''. \textit{Physica B: Condensed Matter} 280, no. 1-4: 20--26.

\bibitem[Griffin 1996]{Griffin96} Griffin, Allan. 1996. ``Conserving and gapless approximations for an inhomogeneous Bose gas at finite temperatures''. \textit{Physical Review B} 53, no. 14: 9341--9347.

\bibitem[Griffiths and Schroeter 2018]{Grif18} Griffiths, David J., and Darrell F. Schroeter. 2018. \textit{Introduction to Quantum Mechanics, Second Edition}. Cambridge, UK: Cambridge University Press.

\bibitem[Gr{\o}nbech-Jensen, Kivshar and Samuelsen 1993]{Gronbech93} Gr{\o}nbech-Jensen, Niels, Yuri S. Kivshar, and Mogens R. Samuelsen. 1993. ``Nonlinear dynamics of a parametrically driven sine-Gordon system''. \textit{Physical Review B} 47, no. 9: 5013--3021.

\bibitem[Grundland and Infeld 1992]{Grundland92} Grundland, A. M., and Eryk Infeld. 1992. ``A family of nonlinear Klein--Gordon equations and their solutions''. \textit{Journal of Mathematical Physics} 33, no. 7: 2498--2503. 

\bibitem[Gross 1961]{Gross61} Gross, Eugene P. 1961. ``Structure of a quantized vortex in boson systems''. \textit{Il Nuovo Cimento (1955-1965) (The New Cimento)} 20, no. 3: 454--477.

\bibitem[Hartree 1928]{Hartree28} Hartree, Douglas R. 1928. ``The wave mechanics of an atom with a non-Coulomb central field. Part I. Theory and methods''. \textit{Mathematical Proceedings of the Cambridge Philosophical Society} 24, no. 1: 89--110.

\bibitem[Hasegawa and Tappert 1973a]{Hasegawa73a} Hasegawa, Akira, and Frederick Tappert. 1973. ``Transmission of stationary nonlinear optical pulses in dispersive dielectric fibers. I. Anomalous dispersion''. \textit{Applied Physics Letters} 23, no. 3: 142--144.

\bibitem[Hasegawa and Tappert 1973b]{Hasegawa73b} Hasegawa, Akira, and Frederick Tappert. 1973. ``Transmission of stationary nonlinear optical pulses in dispersive dielectric fibers. II. Normal dispersion''. \textit{Applied Physics Letters} 23, no. 4: 171--172.

\bibitem[Hasegawa and Matsumoto 2003]{Hasegawa03} Hasegawa, Akira, and Masayuki Matsumoto. 2003. \textit{Optical Solitons in Fibers}. Berlin Heidelberg, Germany: Springer-Verlag.

\bibitem[Headley and Agrawal 2005]{Headley05} Headley, Clifford, and Govind Agrawal. 2005. \textit{Raman Amplification in Fiber Optical Communication Systems}. Boston, MA: Academic Press.

\bibitem[Henderson, Peregrine and Dold 1999]{Henderson99} Henderson, Karen L., D. Howell Peregrine, and John W. Dold. 1999. ``Unsteady water wave modulations: fully nonlinear solutions and comparison with the nonlinear Schr\"odinger equation''. \textit{Wave Motion} 29, no. 4: 341--361.

\bibitem[Infeld and Rowlands 2000, 100--102]{Infeld00} Infeld, Eryk, and George Rowlands. 2000 \textit{Nonlinear Waves, Solitons and Chaos, Second Edition}. Cambridge, UK: Cambridge University Press.

\bibitem[Ivancevic 2010]{Ivancevic10} Ivancevic, Vladimir G. 2010. ``Adaptive-wave alternative for the Black-Scholes option pricing model''. \textit{Cognitive Computation} 2, no. 1: 17--30.

\bibitem[Jim\'enez and V\'azques 1990]{Jimenez90} Jim\'enez, Salvador, and Luis V\'azquez. 1990. ``Analysis of four numerical schemes for a nonlinear Klein-Gordon equation''. \textit{Applied Mathematics and Computation} 35, no. 1: 61--94.

\bibitem[Jin et al 1996]{Jin96} Jin, D. S., J. R. Ensher, M. R. Matthews, C. E. Wieman, and Eric A. Cornell. 1996. ``Collective excitations of a Bose-Einstein condensate in a dilute gas''. \textit{Physical Review Letters} 77, no. 3: 420--423.

\bibitem[Josephson 1962]{Josephson62} Josephson, Brian David. 1962. ``Possible new effects in superconductive tunnelling''.  \textit{Physics Letters} 1, no. 7: 251--253.

\bibitem[Josephson 1974]{Josephson74} Josephson, Brian David. 1974. ``The discovery of tunnelling supercurrents''. \textit{Reviews of Modern Physics} 46, no. 2: 251--254.

\bibitem[Karjanto 2006]{Karjanto06} Karjanto, Natanael. 2006. ``Mathematical Aspects of Extreme Water Waves''. PhD diss., University of Twente, the Netherlands. 

\bibitem[Karjanto and van Groesen 2007a]{Karjanto07a} Karjanto, Natanael, and Embrecht W. C. van Groesen. 2007a. ``Note on wavefront dislocation in surface water waves''. \textit{Physics Letters A} 371: 173--179.

\bibitem[Karjanto and van Groesen 2007b]{Karjanto07b} Karjanto, Natanael, and Embrecht W. C. van Groesen. 2007b. ``Derivation of the NLS breather solutions using displaced phase-amplitude variables''. In \textit{Proceedings of SEAMS-GMU Conference 2007, Section: Applied Mathematics}, pages 357--368. arXiv:1110.4704 [physics.flu-dyn]

\bibitem[Karjanto 2009]{Karjanto09} Karjanto, Natanael, and Embrecht W. C. van Groesen. 2009. ``Mathematical physics properties of waves on finite background''. In Lang, S. P., and Salim H. Bedore. (Eds.) \textit{Handbook of Solitons: Research, Technology and Applications}, pages 501--539. Hauppauge, NY: Nova Science Publishers.

\bibitem[Karjanto and van Groesen 2010]{Karjanto10} Karjanto, Natanael, and Embrecht W. C. van Groesen. 2010. ``Qualitative comparisons of experimental results on deterministic freak wave generation based on modulational instability''. \textit{Journal of Hydro-Environment Research} 3 no. 4: 186--192.

\bibitem[Karpman 1967]{Karpman67} Karpman, Vladimir Iosifovich. 1967. ``Modified conservation laws for nonlinear waves''. \textit{JETP Letters} 6: 277--279.

\bibitem[Karpman and Krushkal 1969]{Karpman69} Karpman, Vladimir Iosifovich, and E. M. Krushkal. 1969. ``Modulated waves in nonlinear dispersive media''. \textit{Soviet Journal of Experimental and Theoretical Physics} 28: 277--281.

\bibitem[Kaup and Newell 1978]{Kaup78} Kaup, David James, and Alan C. Newell. 1978. ``Theory of nonlinear oscillating dipolar excitations in one-dimensional condensates''. \textit{Physical Review B} 18, no. 10: 5162--5167.

\bibitem[Kevorkian and Cole 2012]{Kevorkian12} Kevorkian, Jirair K., and Julian D. Cole. 2012. \textit{Multiple Scale and Singular Perturbation Methods}. Vol. 114. Berlin Heildelberg, Germany: Springer Science \& Business Media.

\bibitem[Kevorkian and Cole 2013]{Kevorkian13} Kevorkian, Jirair K., and Julian D. Cole. 2013. \textit{Perturbation Methods in Applied Mathematics}. Vol. 34. Berlin Heildelberg, Germany: Springer Science \& Business Media.

\bibitem[Kevrekidis, Frantzeskakis and Carretero-Gonz\'ales 2008]{Kevrekidis08} Kevrekidis, Panayotis G., Dimitri J. Frantzeskakis and Ricardo Carretero-Gonz\'ales. 2008. \textit{Emergent Nonlinear Phenomena in Bose-Einstein Condensates--Theory and Experiment}. Berlin Heidelberg, Germany: Springer-Verlag.

\bibitem[Kharif and Pelinovsky 2003]{Kharif03} Kharif, Christian, and Efim Pelinovsky. 2003. ``Physical mechanisms of the rogue wave phenomenon''. \textit{European Journal of Mechanics-B/Fluids} 22, no. 6: 603--634.

\bibitem[Kharif, Pelinovsky and Slunyaev 2009]{Kharif09} Kharif, Christian, Efim Pelinovsky, and Alexey Slunyaev. 2009. \textit{Rogue Waves in the Ocean}. Advances in Geophysical and Environmental Mechanics and Mathematics. Berlin Heidelberg, Germany: Springer-Verlag.

\bibitem[Kibler et al. 2012]{Kibler12} Kibler, Bertrand, Julien Fatome, Christophe Finot, Guy Millot, Go\"{e}ry Genty, Benjamin Wetzel, Nail Akhmediev, Fr\'{e}d\'{e}ric Dias, and John M. Dudley. 2012. ``Observation of Kuznetsov-Ma soliton dynamics in optical fibre'' \textit{Scientific Reports} 2: 463.

\bibitem[Kivshar 2003]{Kivshar03} Kivshar, Yuri S., and Govind Agrawal. 2003. \textit{Optical Solitons--From Fibers to Photonic Crystals}. San Diego, CA: Academic Press.

\bibitem[Kr\"{a}mer 2013]{Kramer13} Kr\"amer, Patrick. 2013. ``The method of multiple scales for nonlinear Klein-Gordon and Schr\"odinger equations''. Diploma thes., Karlsruhe Institute of Technology, Germany.

\bibitem[Kuchment 2008]{Kuchment08} Kuchment, Peter. 2008. ``Quantum graphs: An introduction and a brief survey''. In Exner, Pavel, Jonathan P. Keating, Peter Kuchment, Toshikazu Sunada, and Alexander Teplyaev. (Eds.) \textit{Analysis on Graphs and its Applications}, pages 291--314. Proceedings of Symposia in Pure Mathematics. Providence, RI: American Mathematical Society (AMS).

\bibitem[Kuznetsov 1977]{Kuznetsov77} Kuznetsov, Evgenii Aleksandrovich. 1977. ``Solitons in a parametrically unstable plasma''. \textit{Doklady Akademii Nauk SSSR (Proceedings of the USSR Academy of Sciences)}, 236: 575--577.

\bibitem[Lake et al 1977]{Lake77} Lake, Bruce M., Henry C. Yuen, Harald Rungaldier, and Warren E. Ferguson. 1977. ``Nonlinear deep-water waves: theory and experiment. Part 2. Evolution of a continuous wave train''. \textit{Journal of Fluid Mechanics} 83, no. 1: 49--74.

\bibitem[Li and Vu-Quoc 1995]{Li95} Li, Shaofan, and Loc Vu-Quoc. 1995. ``Finite difference calculus invariant structure of a class of algorithms for the nonlinear Klein--Gordon equation''. \textit{SIAM Journal on Numerical Analysis} 32, no. 6: 1839--1875.

\bibitem[Ma 1979]{Ma79} Ma, Yan-Chow. 1979. ``The perturbed plane-wave solutions of the cubic Schr\"{o}dinger equation''. \textit{Studies in Applied Mathematics} 60, no. 1: 43--58.

\bibitem[Ma- delung 1927]{Madelung27} Madelung, Erwin. 1927. ``Quantentheorie in hydrodynamischer Form'' (``Quantum theory in hydrodynamic form''). \textit{Zeitschrift f\"ur Physik (Journal of Physics)} 40, no. 3: 322--2326.

\bibitem[Mei 1983]{Mei83} Mei, Chiang C. 1983. \textit{The Applied Dynamics of Ocean Surface Waves}. New York, NY: John Wiley \& Sons.

\bibitem[Modugno et al 2001]{Modugno01} Modugno, Giovanni, Gabriele Ferrari, Giacomo Roati, Robert J. Brecha, A. Simoni, and Massimo Inguscio. 2001. ``Bose-Einstein condensation of potassium atoms by sympathetic cooling''. \textit{Science} 294, no. 5545: 1320--1322.

\bibitem[Mollenauer, Stolen and Gordon 1980]{Molleanauer80} Mollenauer, Linn F., Roger H. Stolen, and James P. Gordon. 1980. ``Experimental observation of picosecond pulse narrowing and solitons in optical fibers''. \textit{Physical Review Letters} 45, no. 13: 1095--1098.

\bibitem[Mollenauer and Gordon 2006]{Mollenauer06} Mollenauer, Linn F., and James P. Gordon. 2006. \textit{Solitons in Optical Fibers: Fundamentals and Applications}. Elsevier, the Netherlands: Elsevier and Boston, MA: Academic Press.

\bibitem[Moloney and Newell 2019]{Moloney19} Moloney, Jerome V., and Alan C. Newell. 2019. \textit{Nonlinear Optics}. Boca Raton, FL: CRC Press.

\bibitem[Nayfeh 2008]{Nayfeh08} Nayfeh, Ali H. 2008. \textit{Perturbation Methods}, Hoboken, NJ: John Wiley \& Sons.

\bibitem[Newell 1985]{Newell85} Newell, Alan C. 1985. \textit{Solitons in Mathematics and Physics}. Vol. 48 of \textit{CBMS-NSF Regional Conference Series in Applied Mathematics}.  Philadelphia, PA: Society for Industrial and Applied Mathematics (SIAM).

\bibitem[Onorato et al 2001]{Onorato01} Onorato, Miguel, Alfred Richard Osborne, Marina Serio, and Serena Bertone. 2001. ``Freak waves in random oceanic sea states''. \textit{Physical Review Letters} 86, no. 25: 5831--5834.


\bibitem[Osborne 2001]{Osborne01} Osborne, Alfred Richard. 2001. ``The random and deterministic dynamics of `rogue waves' in unidirectional, deep-water wave trains''. \textit{Marine Structures} 14, no. 3: 275--293.

\bibitem[Pal 2010]{Pal10} Pal, Bishnu P. 2010. \textit{Guided Wave Optical Components and Devices: Basics, Technology, and Applications}. Boston, MA: Academic Press.

\bibitem[Parkins and Walls 1998]{Parkins98} Parkins, Andrew Scott, and Daniel F. Walls. 1998. ``The physics of trapped dilute-gas Bose--Einstein condensates''. \textit{Physics Reports} 303, no. 1: 1--80.

\bibitem[Parmentier 1993]{Parmentier93} Parmentier R. D. 1993. ``Solitons and long Josephson junctions''. In Weinstock H., and Ralston R.W. (Eds.) \textit{The New Superconducting Electronics}, pages 221--248. NATO ASI Series (Series E: Applied Sciences). Vol 251. Dordrecht, the Netherlands: Springer.

\bibitem[Pelinovsky and Kharif 2016]{Pelinovsky16} Pelinovsky, Efim, and Christian Kharif. (Eds.) 2016. \textit{Extreme Ocean Waves, Second Edition}. Cham, Switzerland: Springer.

\bibitem[Peregrine 1983]{Peregrine83} Peregrine, D. Howell. 1983. ``Water waves, nonlinear Schr\"{o}dinger equations and their solutions''. \textit{The ANZIAM Journal} 25, no. 1: 16--43.

\bibitem[P\'erez-Garc\'ia et al 1996]{Perez96} P\'erez-Garc\'ia, Victor M., Humberto Michinel, J. I. Cirac, M. Lewenstein, and P. Zoller. 1996. ``Low energy excitations of a Bose-Einstein condensate: A time-dependent variational analysis''. \textit{Physical Review Letters} 77, no. 27: 5320.

\bibitem[P\'erez-Garc\'ia et al 1997]{Perez97} P\'erez-Garc\'ia, Victor M., Humberto Michinel, J. I. Cirac, M. Lewenstein, and P. Zoller. 1997. ``Dynamics of Bose-Einstein condensates: Variational solutions of the Gross-Pitaevskii equations''. \textit{Physical Review A} 56, no. 2: 1424--1432.

\bibitem[Perring and Skyrme 1962]{Perring62} Perring, J. K., and T. H. R. Skyrme. 1962. ``A model unified field equation''. \textit{Nuclear Physics} 31: 550--555.

\bibitem[Pethick and Smith 2008]{Pethick08} Pethick, Christopher J., and Henrik Smith. 2008. \textit{Bose-Einstein Condensation in Dilute Gases, Second Edition}. Cambridge, UK: Cambridge University Press.

\bibitem[Phillips 2003]{Phil03} Phillips, Anthony C. 2003. \textit{Introduction to Quantum Mechanics}. Hoboken, NJ: John Wiley \& Sons.

\bibitem[Pitaevskii 1961]{Pitaevskii61} Pitaevskii, Lev P. 1961. ``Vortex lines in an imperfect Bose gas''. \textit{Soviet Physics JETP} 13, no. 2: 451--454.

\bibitem[Pitaevskii and Stringari 2003]{Pitaevskii03} Pitaevskii, Lev P., and Sandro Stringari. 2003. \textit{Bose-Einstein Condensation}. Oxford, UK: Clarendon Press.

\bibitem[Pnevmatikos and Pedersen 1993]{Pnevmatikos93} Pnevmatikos Stephanos, and Niels F. Pedersen. 1993. ``The sine-Gordon equation and superconducting soliton oscillators''. In Christiansen P.L., Eilbeck J.C., Parmentier R.D. (Eds.) \textit{Future Directions of Nonlinear Dynamics in Physical and Biological Systems}, pages 283--331. NATO ASI Series (Series B: Physics). Vol 312. Boston, MA: Springer.

\bibitem[Porsezian and Kuriakose 2003]{Porsezian03} Porsezian, Kuppuswamy, and Valakkattil Chako Kuriakose, Eds. 2003. \textit{Optical Solitons: Theoretical and Experimental Challenges}. Vol. 613. Berlin Heildelberg, Germany: Springer-Verlag.

\bibitem[Robert et al 2001]{Robert01} Robert, Alice, Olivier Sirjean, Antoine Browaeys, Julie Poupard, Stephan Nowak, Denis Boiron, Christoph I. Westbrook, and Alain Aspect. 2001. ``A Bose-Einstein condensate of metastable atoms''. \textit{Science} 292, no. 5516: 461--464.

\bibitem[Rogel-Salazar 2013]{Roger13} Rogel-Salazar, Jesus. 2013. ``The Gross-Pitaevskii equation and Bose-Einstein condensates''. \textit{European Journal of Physics} 34, no. 2: 247--257.

\bibitem[Ruprecht et al 1995]{Ruprecht95} Ruprecht, P. A, M. J. Holland, K. Burnett, and Mark Edwards. 1995. ``Time-dependent solution of the nonlinear Schr\"odinger equation for Bose-condensed trapped neutral atoms''. \textit{Physical Review} A 51, no. 6: 4704--4711.

\bibitem[Ruprecht et al 1996]{Ruprecht96} Ruprecht, P. A., Mark Edwards, K. Burnett, and Charles W. Clark. 1996. ``Probing the linear and nonlinear excitations of Bose-condensed neutral atoms in a trap''. \textit{Physical Review A} 54, no. 5: 4178--4187.

\bibitem[Schneider 2011]{Schneider11} Schneider, Guido. 2011. ``The role of the nonlinear Schr\"odinger equation in nonlinear optics''. In D\"{o}rfler, Willy, Armin Lechleiter, Michael Plum, Guido Schneider, and Christian Wieners. (Eds.) \textit{Photonic Crystals: Mathematical Analysis and Numerical Approximation}, pages 127--162. Vol. 42. Berlin/Heidelberg, Germany: Springer Science \& Business Media.

\bibitem[Schr\"odinger 1926]{Schrodinger26} Schr\"odinger, Erwin. 1926. ``An undulatory theory of the mechanics of atoms and molecules''. \textit{The Physical Review} 28, no. 6: 1049--1070.

\bibitem[Shankar 1994]{Shankar94} Shankar, Ramamurti. 1994. \textit{Principles of Quantum Mechanics, Second Edition}. Berlin Heidelberg, Germany: Springer Science \& Business Media.

\bibitem[Sharma and Buti 1976]{Sharma76} Sharma, A. S., and B. Buti. 1976. ``Envelope solitons and holes for sine-Gordon and non-linear Klein-Gordon equations''. \textit{Journal of Physics A: Mathematical and General} 9, no. 11: 1823--1826.

\bibitem[Solli et al 2007]{Solli07} Solli, Daniel R., Claus Ropers, P. Koonath, and Bahram Jalali. 2007. ``Optical rogue waves''. \textit{Nature} 450, no. 7172: 1054--1057.

\bibitem[Stenflo and Marklund 2010]{Stenflo10} Stenflo, Lennart, and Mattias Marklund. 2010. ``Rogue waves in the atmosphere''. \textit{Journal of Plasma Physics} 76, no. 3--4: 293--295.

\bibitem[Strauss and V\'azquez 1978]{Strauss78} Strauss, Walter, and Luis V\'azquez. 1978. ``Numerical solution of a nonlinear Klein-Gordon equation''. \textit{Journal of Computational Physics} 28, no. 2: 271--278.

\bibitem[Strauss 1989]{Strauss89} Strauss, Walter A. 1989. \textit{Nonlinear Wave Equations}. Providence, RI: American Mathematical Society.

\bibitem[Stringari 1996]{Stringari96} Stringari, Sandro. 1996. ``Collective excitations of a trapped Bose-condensed gas''. \textit{Physical Review Letters} 77, no. 12: 2360--2363.

\bibitem[Sulem and Sulem 1999]{Sulem99} Sulem, Catherine and Pierre-Louis Sulem. 1999. \textit{The Nonlinear Schr\"{o}dinger Equation--Self-Focusing and Wave Collapse}. New York, NY: Springer-Verlag.

\bibitem[Susanto et al 2019]{Susanto19} Susanto, Hadi, Natanael Karjanto, Zulkarnain, Toto Nusantara, and Taufiq Widjanarko. 2019. ``Soliton and breather splitting on star graphs from tricrystal Josephson junctions''. \textit{Symmetry} 11, no. 2: 271.

\bibitem[Teschl 2014]{Teschl14} Teschl, Gerald. 2014. \textit{Mathematical Methods in Quantum Mechanics with Applications to Schr\"{o}dinger Operators, Second Edition}. Providence, RI: American Mathematical Society.

\bibitem[Thomas 1927]{Thomas27} Thomas, Llewellyn Hilleth. 1927. ``The calculation of atomic fields''. \textit{Mathematical Proceedings of the Cambridge Philosophical Society} 23, no. 5: 542--548.

\bibitem[Trillo and Torruellas 2001]{Trillo01} Trillo, Stefano, and William Torruellas. (Eds.) 2001. \textit{Spatial Solitons}. Vol. 82. Berlin Heildelberg, Germany: Springer.

\bibitem[van Groesen 1998]{vanGroesen98} van Groesen, Embrecht W. C. 1998. ``Wave groups in uni-directional surface-wave model''. \textit{Journal of Engineering Mathematics} 34, no. 1-2: 215--226.

\bibitem[van Groesen, Andonowati and Karjanto 2006]{vanGroesen06} van Groesen, Embrecht, W. C., Andonowati, and Natanael Karjanto. 2006. ``Displaced phase-amplitude variables for waves on finite background''. \textit{Physics Letters A} 354, no. 4: 312--319.

\bibitem[von Neumann 1932]{von32} Von Neumann, John. 1932. \textit{Mathematical Foundations of Quantum Mechanics} Princeton, NJ and Oxford, UK: Princeton University Press. (The new edition was published in 2018.)

\bibitem[Vu-Quoc and Li 1993]{Vu93} Vu-Quoc, Loc, and Shaofan Li. 1993. ``Invariant-conserving finite difference algorithms for the nonlinear Klein-Gordon equation''. \textit{Computer Methods in Applied Mechanics and Engineering} 107, no. 3: 341--391.

\bibitem[Wazwaz 2005]{Wazwaz05} Wazwaz, Abdul-Majid. 2005. ``The tanh and the sine--cosine methods for compact and noncompact solutions of the nonlinear Klein--Gordon equation''. \textit{Applied Mathematics and Computation} 167, no. 2: 1179--1195.

\bibitem[Weber et al 2003]{Weber03} Weber, Tino, Jens Herbig, Michael Mark, Hanns-Christoph N\"agerl, and Rudolf Grimm. 2003. ``Bose-Einstein condensation of cesium''. \textit{Science} 299, no. 5604: 232--235.

\bibitem[Yan 2010]{Yan10} Yan, Zhen-Ya. 2010. ``Financial rogue waves''. \textit{Communications in Theoretical Physics} 54, no. 5: 947--949.

\bibitem[Zakharov and Shabat 1972]{Zakharov72} Zakharov, V. E., and A. B. Shabat. 1972. ``Exact theory of two-dimensional self-focusing and one-dimensional self-modulation of waves in nonlinear media''. \textit{Soviet Physics JETP} 34, no. 1: 62--69.
\end{thebibliography}
\end{document}